\documentclass[12pt,a4paper]{article}
\usepackage{ifpdf}
\ifpdf
\pdfpageattr {/Group << /S /Transparency /I true /CS /DeviceRGB>>}
\fi
\usepackage{jheppub}
\usepackage{amsmath}
\usepackage{caption}
\usepackage{subcaption}
\usepackage{youngtab}

\newcommand{\cW}{\mathcal{W}}
\newcommand{\cS}{\mathcal{S}}

\newcommand{\SU}{\mathrm{SU}}

\def\UU{\mathrm{U}}
\def\bZ{\mathbb{Z}}
\def\bR{\mathbb{R}}
\def\bC{\mathbb{C}}
\def\cN{\mathcal{N}}
\def\tr{\mathop{\mathrm{tr}}}
\def\diag{\mathop{\mathrm{diag}}}

\newcommand{\beq}{\begin{equation}}
\newcommand{\eeq}{\end{equation}}
\newcommand{\beqa}{\begin{eqnarray}}
\newcommand{\eeqa}{\end{eqnarray}}
\newcommand{\dd}{{\rm d}}

\newcommand{\ZZ}{{\mathbb Z}}
\newcommand{\Z}{\ZZ}

\newcommand{\RR}{{\mathbb R}}
\newcommand{\R}{\RR}
\newcommand{\CC}{{\mathbb C}}
\newcommand{\C}{\CC}

\newcommand{\e}{\,{\rm e}}

\newcommand{\nn}{\nonumber}

\newcommand{\bpsi}{\overline{\psi}}

\newcommand{\blambda}{\overline{\lambda}}

\def\to{\rightarrow}

\begin{document}

\title{2d SCFTs from M2-branes}
\author[a]{Kentaro Hori,}
\author[b]{Chan Y. Park,}
\author[a,c]{and Yuji Tachikawa}
\affiliation[a]{Institute for the Physics and Mathematics of the Universe, \\
  University of Tokyo, Kashiwa, Chiba 277-8583, Japan}
\affiliation[b]{California Institute of Technology, \\
  Pasadena, CA 91125, USA }
\affiliation[c]{Department of Physics, Faculty of Science, \\
  University of Tokyo, Bunkyo-ku, Tokyo 133-0022, Japan}
\preprint{IPMU-13-0177, CALT-68-2858, UT-13-32}  
\abstract{We consider
  the low-energy limit of the two-dimensional theory on $k$ M2-branes
  suspended between a straight M5-brane and a curved M5-brane.  We
  argue that it is described by  an $\cN{=}(2,2)$ supersymmetric gauge theory
with no matter fields but with a non-trivial twisted superpotential,
and also by  an $\cN{=}(2,2)$ supersymmetric Landau-Ginzburg model, such that
the (twisted) superpotentials are determined by the shape of the M5-branes.
 We find  particular cases realize Kazama-Suzuki models. Evidence is provided
  by the study of ground states, chiral rings,
  BPS spectra and $S^2$ partition functions of the
  systems. }

\setcounter{tocdepth}{2}
\maketitle

\section{Introduction and Summary}\label{introduction}

\begin{figure}
\centering
\[
\begin{array}{l||c|c|c|c|c|c|c|c|c|c}
&x^0&x^1&x^2&x^3&v&x^6&t&x^8&x^9 \\
\hline 
\text{M5}& - & - & - & - & - & L & t(v) & 0 & 0  \\
\text{M5$'$}& - & - & 0 & 0 & - & 0 & 1 & - & - \\
\text{M2}& - & - & 0 & 0 & \sigma & - & 1 & 0 & 0
\end{array}\qquad
\vcenter{\hbox{\includegraphics[width=.4\textwidth]{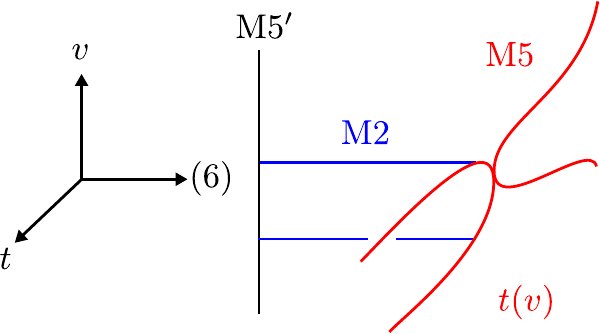}}}
\]

\caption{Configuration of branes. $v=x^4+ix^5$ and $t=\exp(x^7+ix^{10})$.
\label{config}}
\end{figure}

Consideration of multiple $p$-branes suspended between other branes is 
an effective way to study the dynamics of $p$-dimensional supersymmetric
field theories \cite{Hanany:1996ie,Witten:1997sc}. 
In this paper, we  analyze the dynamics of multiple
M2-branes suspended between two M5-branes in the following setup
\cite{Hanany:1997vm}. 
We use the coordinates $x^{0,\cdots,10}$, with a compactified $x^{10}$ direction.
Let us introduce complex combinations $v=x^4+ix^5$ and $t=\exp(x^7+ix^{10})$. 
Then we have, as summarized in Figure \ref{config}, \begin{itemize}
\item an M5-brane  extending along $x^{0,1,2,3}$ and on the complex
one-dimensional curve $t=t(v)$,  at a fixed position 
$(x^6,x^8,x^9)=(L,0,0)$,
\item an M5-brane (which we call the M5$'$-brane) extending along
directions $x^{0,1,8,9}$ and $v$, at a fixed potition 
$(x^2,x^3,x^6,t)=(0,0,0,1)$, and
\item $k$ M2-branes extending along $x^{0,1}$ and suspended between
the M5 and the M5$'$ along the $x^6$ direction.
\end{itemize}
We are interested in the infrared limit of the theory on $k$ M2-branes.

If we reduce the theory along the $x^{10}$ direction,
we have a system of $k$ D2-branes suspended between one NS5-brane at $x^6=0$
and some configuration of branes at $x^6=L$.
This gives a 3d $\UU(k)$ gauge theory formulated on an interval
with some boundary conditions at the two ends,
which reduces to a 2d theory with $\cN{=}(2,2)$ supersymmetry
at distances longer than the length $L$ of the interval.
We first assume that all solutions to the equation $t(v)=1$
are non-degenerate. That is, if $\{v_i\}_{i\in I}$ denotes
the set of solutions, then $t'(v_i)\ne 0$ for each $i\in I$.
Then, from the M-theory description, the supersymmetric
vacua are described by $k$ M2-branes, separated along $v$ directions,
each at a fixed value of $v$ being one of $\{v_i\}_{i\in I}$.
The s-rule \cite{Hanany:1996ie, Hanany:1997vm} forbids that more than one
M2-brane have the same value of $v$.
This vacuum structure would arize if the low energy theory is
the theory on the Coulomb branch with the twisted superpotential
\begin{equation}
	\cW_{\rm eff}=\tr P(\Sigma_T),\label{twistedsuperpotential}
\end{equation} 
for the fieldstrength superfield
$\Sigma_T={\rm diag}(\Sigma_1,\ldots,\Sigma_k)$ for the maximal torus
$T\cong\UU(1)^k$,
where the holomorphic function $P(v)$ is given by 
\begin{equation}
\exp(P'(v))=t(v).\label{definitionofP}
\end{equation}
Indeed, the vacuum equation is $t(\sigma_a)=1$ for $a=1,\ldots,k$
and we expect that no supersymmetric vacuum is supported at the solutions with
$\sigma_a=\sigma_b$ for $a\neq b$.
Also, permutations of $\sigma_a$'s are gauge symmetry.

When $t(v)$ is a rational function, the M5 reduces to a number of
D4-branes ending on an NS5-brane, and the 2d theory can be interpreted as
a $\UU(k)$ gauge theory with a number of fundamental and antifundamental
chiral multiplets, possibly with twisted masses \cite{Hanany:1997vm}. 
There are only finitely many solutions to $t(v)=1$
and hence the number of supersymmetric vacua is finite.
When $t(v)$ is such that $P'(v)$ in (\ref{definitionofP})
is a polynomial, the 2d theory has a different type of interpretation:
It is the $\UU(k)$ gauge theory without matter field and with the tree level
twisted superpotential
\beq
\cW=\tr P(\Sigma)+\pi i(k+1)\tr\Sigma,
\label{Wtree}
\eeq
where $\Sigma$ is now the fieldstrength for the full $\UU(k)$ vector
mutiplet. The second term is the theta term with $\theta=\pi (k+1)$.
It is non-trivial if and only if $k$ is even since
$\theta$ is a periodic parameter of period $2\pi$.
This is needed in order to have (\ref{twistedsuperpotential})
as the effective twisted superpotential on the Coulomb branch \cite{Hori:2011pd}.
The equation $t(v)=1$ has an infinitely many solutions, and correspondingly,
there are infinitely many supersymmetric vacua in this gauge system.

Each vacuum has a mass gap when, as assumed above, all the solutions to
$t(v)=1$ are non-degenerate.
Things would be more interesting if $t(v)$ is fine tuned
so that some of the solutions coincide, or equivalently,
some of the solutions are degenerate.
$N$ solutions coincide, say at $v=0$, when
\beq
t(v)=1+v^N+\cdots\quad\mbox{or}\quad
P(v)=v^{N+1}+\cdots,
\label{Pv}
\eeq
where the ellipses stand for possible terms of higher order in $v$.
In such a case, we expect to have a non-trivial conformal field theory
in the infra-red limit. In fact, for the case $k=1$, it is argued
in \cite{Tong:2006pa} that the vacuum at $v=0$ is the same as
the infra-red limit of the Landau-Ginzburg model with superpotential
$W=X^{N+1}$, which is believed to be equivalent to
the $\cN{=}(2,2)$ superconformal minimal model of type $A_{N-1}$.
For $k>1$, we may have vacua where multiple M2 branes are at $v=0$.
We expect that all $k$ of them can sit there as long as $N\geq k$.
We would like to ask: What is the infra-red limit of such a theory?

We will argue that the theory under question is equivalent
to the infra-red limit of
the Landau-Ginzburg model of $k$ variables $X_1,\ldots, X_k$,
where the superpotential $W(X_1,\ldots,X_k)$ is $\tr \Sigma_T^{N+1}$
written in terms of the elementary symmetric functions of
$\Sigma_1,\ldots,\Sigma_k$;
\beqa
\sum_{a=1}^k\sigma_a^{N+1}&=&W(x_1,\ldots,x_k),
\label{WP}\\
x_b&=&\sum_{a_1<\cdots<a_b}\sigma_{a_1}\cdots\sigma_{a_b},
\qquad b=1,\ldots,k
\eeqa
This model is believed to flow to the $\cN{=}(2,2)$ superconformal
Kazama-Suzuki model \cite{Kazama:1988qp,Kazama:1988uz}
of the coset type
\begin{equation} \frac{\SU(N)_1}{\mathrm{S}[\UU(k)\times
    \UU(N-k)]}.\label{KS}
\end{equation}
Thus, we claim that the answer to the question is this Kazama-Suzuki
model.

The behaviour (\ref{Pv}) is realized simply by 
$t(v)=1+v^N$ or $P(v)=v^{N+1}$.
In the former case, the 2d theory is the $\UU(k)$ SQCD 
with $N$ fundamental matter fields with fine tuned twisted masses
\cite{Hanany:1997vm,Tong:2006pa}.
In the latter case, the 2d theory is the pure $\UU(k)$ gauge theory with
the tree level twisted superpotential
\beq
\cW=\tr \Sigma^{N+1}+\pi i(k+1)\tr\Sigma.
\label{pureW}
\eeq
For $1\leq k\leq N$, we shall argue that the theory has, among infinitely
many others, a set of ground states supported at $\Sigma=0$,
and this ``$\Sigma=0$ sector" flows to the superconformal
field theory under question.

Thus, we have a purely field theoretical duality statement: 
for $1\leq k\leq N$,
\begin{itemize}
\item the $\UU(k)$ SQCD with $N$ fundamentals having fine tuned twisted masses,
\item the $\Sigma=0$ sector of the pure $\UU(k)$ gauge theory with
superpotential (\ref{pureW}),
and
\item  the Landau-Ginzburg model with superpotential (\ref{WP})
\end{itemize}
 all 
flow to the infra-red fixed point given by the Kazama-Suzuki model  (\ref{KS}).

The aim of this paper is to give evidence of the claims above. 
We compute the number of supersymmetric ground states and
the chiral ring in the respective systems,
and show that they agree.
We also study the BPS spectrum of the brane system and compare
it with the known field theoretical results.  Superconformal points
themselves are hard to analyze, and therefore we often make mass
deformations.  We also study the $S^2$ partition functions by using the
recently-developed technique of exact computations
\cite{Benini:2012ui,Doroud:2012xw,Gomis:2012wy,Honda:2013uca,Hori:2013ika}
and show that they indeed agree.

The rest of the paper is organized as follows. 
In Section~\ref{sec:vac} and \ref{sec:chiralring},
we test our claim by studying supersymmetric ground states
 and the chiral ring.
In Section \ref{sec:BPS}, we study BPS solitions of
the M2-brane system and compare the structure with the BPS
spectrum of the Landau-Ginzburg theory \cite{Lerche:1991re}. 
If we take a certain limit, the BPS spectrum can be determined
by using the technique of spectral networks \cite{Gaiotto:2012rg}.
In Section \ref{sec:pf}, we calculate the $S^2$
partition functions of the 2d theories on both sides of the claimed
equivalence and show that they agree.
The sections \ref{sec:BPS} and \ref{sec:pf} can be read independently.  
In Appendix~\ref{KSreview}, we
briefly review the basic facts of Kazama-Suzuki models and their
correspondence with Landau-Ginzburg theories.
In Appendix~\ref{Morse}, we give a proof of some algebraic statement
needed for the study of chiral ring.
In Appendix~\ref{convergence}, we discuss the convergence of the integral
appearing in the $S^2$ partition function.

\section{Supersymmetric Vacua}\label{sec:vac}

As the first check, we look at the supersymmetric vacua of the respective
systems or compute the Witten index \cite{Witten:1982df},
and see if the results are consistent with the claimed duality.

\subsection{Brane System}

Let us first look at the brane system. As in the introduction,
we denote the set of solutions to $t(v)=1$ by $\{v_i\}_{i\in I}$
where we initially assume that each solution is non-degenerate $t'(v_i)=0$.
Supersymmetry requires each M2-brane to have a fixed position in $(t,v)$.
The boundary at M5$'$ fixes $t$ to be $1$ and allowes $v$ to be arbitrary,
while the boundary at M5 requires the relation $t=t(v)$.
Thus, each M2 must be at $t=1$ and has $v=v_i$ for some $i\in I$.
The s-rule requires different M2 branes to have different values of
$v$. Therefore, a supersymmetric vacuum is specified by picking $k$
distinct elements from this set: 
\begin{equation} V\subset
  \{v_i\}_{i\in I}, \quad |V|=k.\label{vacua}
\end{equation}
When $t(v)$ is a polynomial of order $M$, the equation $t(v)=1$ has
$M$ roots. Generically, they are distinct and non-degenerate.
Then, the number of supersymmetric vacua is $0$ if $k>M$ and ${M\choose k}$
if $k\leq M$. We may also consider a special polynomial where some of the
solutions coincide. In this situation we do not know how to identify the
supersymmetric vacua. However, the Witten index \cite{Witten:1982df},
which does not change under continuous deformation, remains the same as
in the non-degenerate case.
When some number, say $N$, of the solutions are close to each other
while others are far away, then, we may consider the ``subsector'' in which
all $k$ M2 branes are at one of these $N$ solutions. 
In particular,
when $N\geq k$ of them are at the same point,
we expect to have a {\it single} infra-red theory whose Witten index is 
${N\choose k}$.
This discussion on subsectors and their Witten indices
is applicable even when $t(v)$ is not a polynomial and the equation
$t(v)=1$ have infinitely many solutions.

\subsection{Gauge Theory}

Let us next consider the $\UU(k)$ gauge theory with the tree level twisted
superpotential (\ref{Wtree}) determined by a polynomial $P(\sigma)$.
The classical scalar potential takes the form
\beq
U={1\over 4g^2}\tr[\sigma,\sigma^{\dag}]^2
+{g^2\over 2}\tr ({\rm Re}P'(\sigma))^2.
\label{Uclass}
\eeq
Vanishing of the first term requires $\sigma$ to be diagonalizable,
\beq
\sigma=\diag(\sigma_1,\ldots,\sigma_k).
\label{Coulomb}
\eeq
When all the eigenvalues are well separated, the value of $\sigma$
breaks the gauge group $\UU(k)$ to its diagonal subgroup $T\cong \UU(1)^k$.
In this Coulomb branch, we may integrate out the off-diagonal components of
the vector multiplet. This induces a correction to the twisted superpotential.
As explained in \cite{Hori:2011pd} following \cite{Witten:1993xi}, 
the correction is given by $\pi i$
times the sum of positive roots,\footnote{Incidentally, this settles
a problem concerning the relation  observed in \cite{Hori:2006dk}
between the tree level theta angle
of the $\UU(k)$ linear sigma model and the B-field of the corresponding
non-linear sigma model on a complete intersection
in the Grassmannian $G(k,N)$.
The shift (\ref{shifttheta}) was missed in \cite{Hori:2006dk}
and the relation is mistakenly stated as
``$B=\theta+(N+k+1)\pi$''. This must be corrected to $B=\theta+N\pi$.
Now it is understandable as result from integrating
out the ``$P$-fields'' \cite{Morrison:1994fr}.}
\beq
\Delta \cW=\pi i\sum_{a<b}(\Sigma_a-\Sigma_b)
\equiv \pi i (k+1)\sum_{a=1}^k\Sigma_a.
\label{shifttheta}
\eeq
In the second equality, we used the periodicity $\theta_a\equiv\theta_a+2\pi$
of the theta angle for the group $T$.
This cancels the tree level theta term in (\ref{Wtree})
and hence the effective twisted superpotential is
\beq
\cW_{\rm eff}=\cW|_{T}+\Delta\cW=\sum_{a=1}^kP(\Sigma_a).
\eeq
We denote the effective gauge coupling constant by $e^2_{ab}(\sigma)$. We know
that it approaches $g^2\delta_{ab}$ in the limit where all $\sigma_a$ are
infinitely separated. 
We assume that it is positive definite in the region of $\sigma$
we are looking at, 
and defines inner products, $\Vert y\Vert^2_{e^{-2}}=(e^{-2})^{ab}y_ay_b$ and
  $\Vert x\Vert^2_{e^2}=e^2_{ab}x^ax^b$, on the Lie agbera of $T$ and its dual.
The effective potential is given by
\beq
U_{\rm eff}={1\over 2}\left\Vert
{\rm Re}\cW'_{\rm eff}(\sigma)\right\Vert^2_{e^2}
+{1\over 2}\left\Vert v_{01}\right\Vert^2_{e^{-2}}.
\label{Ueff}
\eeq
The first term, where
${\rm Re}(\cW'_{\rm eff}(\sigma))^a
={\rm Re}P'(\sigma_a)$,  is the remnant of 
the classical potential (\ref{Uclass}).
The second term is the electro-static energy
\cite{Coleman:1976uz,Witten:1993yc}.
In the Hamiltonian formulation, see e.g.~\cite{Hori:2003ic},
$(e^{-2})^{ab}v_{b01}+{\rm Im}(\cW'_{\rm eff}(\sigma))^a$ are regarded as
the conjugate momenta for the holonomy of $T$, each of which has period $1$,
and hence have eigenvalues in $2\pi\Z$. In other words,
\beq
v_{a01}=\sum_{b=1}^k
e^2_{ab}(\sigma)\left(2\pi n^b-{\rm Im}P'(\sigma_b)\right)
\label{vW}
\eeq
where $n^a\in \Z$. In the sector with definite $n^a$'s,
the effective potential is 
\beq
U_{\rm eff}
=\sum_{a,b=1}^k{e^2_{ab}(\sigma)\over 2}(P'(\sigma_a)-2\pi i n^a)
\overline{(P'(\sigma_b)-2\pi i n^b)}.
\label{Ueff2}
\eeq
Supersymmetric ground states must be at the zeroes of this potential.
That is, each $(\sigma_a,n^a)$ must satisfy
\beq
P'(\sigma)=2\pi i n,\qquad n\in \Z.
\label{vacueq}
\eeq
The above analysis is valid only when $\sigma_1,\ldots,\sigma_k$ 
are separated. We do not know how to analyze the region
near the diagonals where some of $\sigma_a$'s coincide.
In many examples, however, it is found that no supersymmetric ground state is
supported near the diagonals as long as the critical points of
the effective twisted superpotential are all non-degenerate.
See for example \cite{Hori:2006dk}. Here we assume that this
applies to our system. Note also that solutions related by permutations of
$(\sigma_a,n^a)$'s are related by the residual gauge transformations and must
be identified. Thus, when $P''(\sigma)\ne 0$ at each solution
to (\ref{vacueq}), a supersymmetric vacuum is specified by a choice of
$k$ unordered solutions $\{(\sigma_a,n^a)\}$ to (\ref{vacueq})
such that $\sigma_a\ne \sigma_b$ for $a\ne b$.
We see that there are infinitely many supersymmetric vacua.

The equation (\ref{vacueq}) may be written simply as
$\exp(P'(v))=1$. Then we see that the problem of finding supersymmetric
vacua in this system is identical to that in the M2 brane system where
the function $t(v)$ defining the M5 curve is given by (\ref{definitionofP}).

Let us write
\beq
P_u(\sigma)={1\over N+1}\sigma^{N+1}+\sum_{j=1}^N{u_j\over N+1-j}\sigma^{N+1-j}.
\label{Psu}
\eeq
for which the equation (\ref{vacueq}) reads
\beq
\sigma^N+\sum_{j=1}^Nu_j\sigma^{N-j}=2\pi i n,\qquad n\in\Z.
\label{releq}
\eeq
For a small but generic $u = (u_1,\ldots, u_N)$, the equation with
$n=0$ has $N$ distinct solutions close to $\sigma=0$,
while the equation with $n\ne 0$ has $N$ separated solutions at
$|v|\sim (2\pi n)^{1/N}$.
Our main interest will be the sector with $n^1=\cdots=n^k=0$.
The supersymmetric vacua must have $\sigma_a$ values from the
$N$ solutions near $0$.
The number of such vacua 
is zero when $k>N$ and ${N\choose k}$ when $1\leq k\leq N$.
When we turn off $u$, the $N$ solutions all go to $\sigma=0$.
If $1\leq k\leq N$, we expect to have a single infra-red theory 
from the $n^1=\cdots=n^k=0$ sector. Its Witten index is ${N\choose k}$.

\subsection{Landau-Ginzburg Model}

Finally, we consider the Landau-Ginzburg model.
Let $W_u(X)=W_u(X_1,\ldots,X_k)$ be the superpotential corresponding to 
$P_u(\sigma)$ of (\ref{Psu}), that is,
$\sum_{a=1}^kP_u(\Sigma_a)$ written in terms of the elementary symmetric
functions of $\Sigma_1,\ldots,\Sigma_k$.

When we turn off $u$, the superpotential $W_0(X)$ is the one
(\ref{WP}) given in the introduction and is a quasi-homogeneous
polynomial. When $N\geq k$, it has an isolated critical point
at $X=0$ and the Landau-Ginzburg model
is believed to flow to a non-trivial superconformal
field theory of central charge $c=3k(N-k)/(N+1)$.
In fact the conformal field theory has been claimed to be equivalent to 
the Kazama-Sukuki supercoset of the type (\ref{KS}). 
See Appendix~\ref{KSreview}. The space of
supersymmetric ground states of the model is naturally identified with
the representation $\wedge^k\C^N$ of $\SU(N)$ \cite{Lerche:1989uy}.
Its dimension ${N\choose k}$ matches the Witten index
of the M2 and the gauge systems.

The model with $u_j\ne 0$ can be regarded as a perturbation of
this superconformal field theory by the chiral primary fields $\phi_j(X)$
corresponding to $\sum_{a=1}\sigma_a^{N+1-j}$. These have R-charges
$2(N+1-j)/(N+1)$ and conformal weigths $(N+1-j)/(N+1)<1$ and hence
the perturbation is relevant.
In particular, the number of supersymmetric ground states
remains the same, ${N\choose k}$.
Moreover, for the particular deformation
where all $u_j$ but $u_N$ vanish, the ground states are labelled by
 the weights of the representation $\wedge^k\C^N$ of $\SU(N)$ 
mentioned above \cite{Lerche:1991re}.
This picture matches with the one for the M2 and the gauge systems
if we regard the roots of $\sigma^N+u_N=0$
as the weights of the representation $\C^N$.
This observation will be important when we compare the spectra of
BPS solitons.

For a generic choice of $u$, the correspondence of the ground states
with those of the gauge system can be seen more explicitly.
The map $\sigma\mapsto x(\sigma)$, defined by
the elementary symmetric functions
$x_1(\sigma),\ldots,x_k(\sigma)$ of $\sigma_1,\ldots,\sigma_k$,
is regular away from the diagonals, since the Jacobi matrix has determinant
\beq
\det\left({\partial x_b\over \partial\sigma_a}\right)_{1\leq a,b\leq k}
=\prod_{1\leq a<b\leq k}(\sigma_a-\sigma_b).\label{jac}
\eeq
The singular values, i.e., the image of the diagonals, shall be called
{\it the discriminant}.
Let us write $f_u(\sigma_1,\ldots,\sigma_k)=\sum_{a=1}^kP_u(\sigma_a)$.
Then, we have
\beq
f_u(\sigma)=W_u(x(\sigma)).
\eeq
Taking the first derivatives, we obtain
\beq
{\partial f_u\over\partial \sigma_a}(\sigma)=\sum_{b=1}^k
{\partial x_b\over\partial\sigma_a}(\sigma)
{\partial W_u\over\partial x_b}(x(\sigma)).
\label{1stder}
\eeq
This means that ``off the diagonals'' critical points of
$f_u(\sigma)$ modulo permutations of $\sigma_a$'s
are in one-to-one correspondence with ``off the discriminant''
critical points of $W_u(x)$.
Taking one more $\sigma$ derivative
and computing the determinant, one sees that
the Hessian of $f_u(\sigma)$ vanishes if $x(\sigma)$ is
a critical point of $W_u(x)$ on the discriminant.
Therefore, if all the critical points of 
$f_u(\sigma)$ are non-degenerate, then, all the critical points of
$W_u(x)$, if there exist, are off the discriminant
and also non-degenerate.
(Note however that $f_u(\sigma)$ may have a non-degenerate
critical point on the diagonal that does not correspond to a critical point
of $W_u(x)$.) 
This establishes a one-to-one correspondence between the 
supersymmetric ground states of the $n^1=\cdots =n^k=0$ sector of
the gauge system and those of the Landau-Ginzburg model, for a generic
$u$ so that $f_u(\sigma)$ is a Morse function.
In particular, this is one way to see that
the number of critical points of $W_u(X)$ is zero for $N<k$
and ${N\choose k}$ for $N\geq k$.

\section{Chiral Rings}\label{sec:chiralring}

In this section, we shall study the chiral ring of
the gauge system and compare the result with 
that of the Landau-Ginzburg model.
We consider the $\UU(k)$ gauge theory with tree level twisted superpotential
\beq
\cW=f(\Sigma)+\pi i(k+1)\tr\Sigma
\eeq
where $f(\Sigma)$ is an adjoint invariant polynomal of $\Sigma$.
The effective twisted superpotential on the Coulomb branch is
$\cW_{\rm eff}=f(\Sigma_T)$.
We shall use the same notation $f(\sigma)=f(\sigma_1,\ldots,\sigma_k)$
for that symmetric polynomial,
and denote simply by $W(X)$ the corresponding superpotential,
$f(\sigma)=W(x(\sigma))$. Just as in (\ref{1stder}), we have
\beq
{\partial f\over\partial \sigma_a}(\sigma)=\sum_{b=1}^k
{\partial x_b\over\partial\sigma_a}(\sigma)
{\partial W\over\partial x_b}(x(\sigma)).
\label{1st}
\eeq
We assume that $f(\sigma)$ is a Morse function.
Then, $W(x)$ is also Morse, and
supersymmetric ground states of the $n^1=\cdots n^k=0$ sector of
the gauge system are in one-to-one correspondence with
those of the Landau-Ginzburg model.

The chiral ring of the Landau-Ginzburg model is generated by the
chiral variables $x_1,\ldots, x_k$ and the relations are
generated by
\beq
0\equiv \left\{{\bf Q}_B,g^{a\bar b}(\bpsi_{b-}-\bpsi_{b+})\right\}
=\partial_{x_a}W(x),\qquad a=1,\ldots,k.
\eeq
Here ${\bf Q}_B$ is the relevant supercharge,
$g^{a\bar b}$ is the K\"ahler metric that appears in the kinetic term,
and $\bpsi_{b\pm}$ are the fermionic components of the antichiral multiplet
$\overline{X}_b$.
Hence the chiral ring is isomorphic to the Jacobi ring,
\beq
{\rm Jac}(W)=\C[x_1,\ldots, x_k]/(\partial_{x_1}W(x),\ldots,
\partial_{x_k}W(x)).
\eeq

The twisted chiral ring of the gauge system is generated by
gauge invariant polynomials of $\sigma$. In the low energy description
on the Coulomb branch, they reduce to symmetric functions of
$\sigma_1,\ldots, \sigma_k$. To find the relations,
we note that
\beq
\left\{{\bf Q}_A,(e^{-2})^{ab}(\blambda_{b-}-\lambda_{b+})
\right\}=(e^{-2})^{ab}(D_b+iv_{b01})
\label{QAL}
\eeq
where ${\bf Q}_A$ is the relevant supercharge while 
$\blambda_{b-}$, $\lambda_{b+}$ and $D_b$ are fermionic and auxiliary
components of the twisted antichiral multiplet $\overline{\Sigma}_b$.
The auxiliary fields $D_b$ are constrained to be
\beq
(e^{-2})^{ab}D_b=-{\rm Re}\,\partial_{\sigma_a}f(\sigma).
\label{eD}
\eeq
We also have equations like (\ref{vW}):
\beq
(e^{-2})^{ab}v_{b01}=-{\rm Im}\,\partial_{\sigma_a}f(\sigma)+2\pi n^a,
\label{ev}
\eeq
where $n^a$ are integers labeling the momenta of the holonomy variables.
Therefore the relations are
$\partial_{\sigma_a}f(\sigma)\equiv 2\pi in^a$.\footnote{In 
Eqns (\ref{QAL}), (\ref{eD}), (\ref{ev})
 some fermion bilinear terms are ignored to simplify the expression.
However, these final relations are exact.} 
Our main interest is the $n^1=\cdots=n^k=0$ sector. The relations are
\beq
\partial_{\sigma_a}f(\sigma)\equiv 0,\qquad a=1,\ldots, k.
\eeq
We shall also accept relations of the form
\beq
\sum_{a=1}^k{F_a(\sigma)\over \Delta(\sigma)^{\ell}}
\,\partial_{\sigma_a}f(\sigma)\equiv 0
\label{Ifgen}
\eeq
where $F_a(\sigma)$ are polynomials and $\Delta(\sigma)$ is
the Vandermond determinant
\beq
\Delta(\sigma):=\prod_{1\leq a<b\leq k}(\sigma_a-\sigma_b).
\eeq
We allow division by $\Delta(\sigma)$ 
because $\sigma_a$'s are assumed to be separated from each other
in the Coulomb branch. Let $I_f$ be the ideal
of the ring $\C[\sigma_1,\ldots,\sigma_k]^{\mathfrak{S}_k}$ of
symmetric polynomials consisting of polynomials that can be
written in the form on 
 the left hand side of (\ref{Ifgen}).
Then, the twisted chiral ring is
\beq
\C[\sigma_1,\ldots,\sigma_k]^{\mathfrak{S}_k}/I_f.
\eeq
When $f(\sigma)$ is generic so that $W(x)$ has only isolated and 
non-degenerate critical points (i.e. $W(x)$ is Morse),
one can show that this is isomorphic to the Jacobi ring ${\rm Jac}(W)$.

The proof goes as follows. First, we have an isomorphism
$\C[x_1,\ldots,x_k]\cong\C[\sigma_1,\ldots,\sigma_k]^{\mathfrak{S}_k}$
given by $\phi(x)\mapsto \phi(x(\sigma))$. It is enough to show that the
ideal $I_W=(\partial_{x_1}W,\ldots,\partial_{x_k}W)$ is mapped precisely
to $I_f$ under this isomorphism.
That $I_W$ is mapped into $I_f$ is obvious in view of (\ref{1st})
and the definition of $I_f$. 
To show that the map
$I_{W}\to I_f$ is surjective, let $\phi(x)$ be a polynomial so that
$\phi(x(\sigma))$ belongs to $I_f$. Then,
$\phi(x(\sigma))$ vanishes on ``off the diagonals'' critical points of
$f(\sigma)$. Here we recall from the previous section
that $\sigma\mapsto x(\sigma)$ gives one-to-one correspondence between
``off the diagonals'' critical points of $f(\sigma)$ modulo permutations
and critical points of $W(x)$.
Therefore, $\phi(x)$ vanishes on the critical points of $W(x)$.
Since $W(x)$ is a Morse function, this means that $\phi(x)$ belongs to $I_W$.
See Appendix~\ref{Morse} for the proof of the last statement.



\section{BPS Solitons}\label{sec:BPS}

In this section we analyze the spectrum of the BPS states from
M2-branes,
building on \cite{Hanany:1997vm,Fayyazuddin:1997by,
Henningson:1997hy,Mikhailov:1997jv,Dorey:1999zk},
and compare the results with the spectrum of BPS solitons
in the Landau-Ginzburg model \cite{Fendley:1990zj,Lerche:1991re}.

In what follows, we are interested in M2-branes whose $(t,v)$ values
are confined into a small neighborhood of
$v=0$ and $t=1$. Therefore, we write $t=e^z$ and regard $z$ as a coordinate on a neighborhood of the origin of 
of a complex plane $\C$.
M5$'$ is at $z=0$ and 
we consider the M5-brane wrapped on the curve
\begin{equation}
z= v^N +  u_1  v^{N-1} + \cdots +  u_N.\label{curve}
\end{equation}
Recall (\ref{vacua})
 that a supersymmetric ground state is specified for a choice
of $k$ distinct elements from the set $\{v_i\}_{i=1}^N$
of solutions to $v^N +  u_1  v^{N-1} + \cdots +  u_N=0$.
We are interested in solitonic M2-brane configurations that
interpolate two different ground states.

\subsection{A single M2-brane}

Let us recall the basics of BPS solitons arising from a single
M2-brane stretched between  two M5-branes. This setup was originally studied in
\cite{Hanany:1997vm} and later in \cite{Dorey:1999zk}.
The system may be regarded as an $\cN{=}(2,2)$ supersymmetric field theory on
$\R^2=\{(x^0,x^1)\}$ with a chiral multiplet taking values in the space of paths
$\phi:x^6\in [0,L]\mapsto (z(x^6),v(x^6))\in \C^2$
from the M5$'$ at $z=0$ to the M5 at (\ref{curve}). It has
the superpotential
\cite{Witten:1997ep}
\beq
\cW[\phi]=\int_{C_{\phi,\phi_*}}\Omega,\qquad\Omega:=\dd z\wedge \dd v,
\eeq
where $C_{\phi,\phi_*}$ is a configuration that interpolates a reference
path $\phi_*$ and $\phi$.
Note that
\beq
{\delta\cW\over \delta z(x^6)}=\partial_6v(x^6),\qquad
{\delta\cW\over \delta v(x^6)}=-\partial_6 z(x^6).
\eeq
In particular, the action includes the usual kinetic term of a 
 theory on three dimensions $(x^0,x^1,x^6)$.
A soliton is a configuration that approaches two vacua, say
$\phi_j\equiv (0,v_j)$ and $\phi_i\equiv (0,v_i)$,
as $x^1\to-\infty$ and $x^1\to +\infty$ respectively.
The central charge of such a solitonic sector is
\beq
Z_{ij}=\cW[\phi_i]-\cW[\phi_j]
=\int_{C_{\phi_i,\phi_j}}\Omega.
\eeq
A soliton preserves a half of the supersymmetry 
if the configuration satisfies the BPS equation,
$\partial_1\phi=\zeta_{ij}\overline{\delta\cW/\delta\phi}$
with $\zeta_{ij}:=Z_{ij}/|Z_{ij}|$,
i.e.,
\beq
\partial_1z=\zeta_{ij}\partial_6\overline{v},\qquad
\partial_1v=-\zeta_{ij}\partial_6\overline{z}.
\label{BPS}
\eeq
It follows that
\beq
\phi^*\omega=0,\qquad
\overline{\zeta}_{ij}\phi^*\Omega
=\dd x^1\wedge\dd x^6\cdot(\mbox{real positive}),
\eeq
where $\omega:={i\over 2}\dd z\wedge \dd \overline{z}+{i\over 2}\dd v\wedge
\dd \overline{v}$ is the K\"ahler form.
This is equivalent \cite{Henningson:1997hy} to the condition
that the image of $\phi:\R\times [0,L]\to \C^2$
is a special Lagrangian submanifold.

One may also look at the usual
supersymmetry condition \cite{Becker:1995kb,Fayyazuddin:1997by,Dorey:1999zk}.
Let $\eta$ be the eleven-dimensional spinor obeying 
$\eta=\Gamma_{012\ldots 9,10}\eta$.
Presence of the M5-branes imposes the condition
$\eta=\Gamma_{014589}\eta=\Gamma_{012345}\eta=\Gamma_{01789,10}\eta$ from which
we also have $\eta=\Gamma_{016}\eta$.
Then, the BPS equation (\ref{BPS}) is equivalent to
the existence of a spinor $\eta$ obeying
\beq
\eta={1\over 2}
\epsilon^{\alpha\beta}\Gamma_{0IJ}
\partial_{\alpha}x^I\partial_{\beta}x^J\eta
\eeq
(the summation over $I,J=1,4,5,6,7,10$ and $\alpha,\beta=1,6$ is assumed),
in the limit $|\partial_{1,6}x^{4,5,7,10}|\ll 1$ where
the eleven dimensional Planck length is set equal to one.
The preserved supersymmetry is 
$(\zeta_{ij}\Gamma_{vz}+\overline{\zeta}_{ij}\Gamma_{\bar v\bar z})\eta
=\Gamma_{16}\eta$.

If we change the complex structure of $\C^2$ so that
$\omega-i{\rm Im}(\overline{\zeta}_{ij}\Omega)$ is a holomorphic two form
and ${\rm Re}(\overline{\zeta}_{ij}\Omega)$ is a K\"ahler form,
the BPS equation (\ref{BPS}) has a different interpretation:
It is a Cauchy-Riemann equation with respect to the holomorphic coordinate
$x^1+ix^6$ of the domain $\R\times [0,L]$.
That is, $\phi:\R\times [0,L]\to \C^2$ can be regarded as 
a holomorphic map. With respect to the new K\"ahler form, the two M5 curves,
$z=0$ and (\ref{curve}), are Lagrangian submanifolds of $\C^2$
which intersects at the $N$ points $\{(0,v_i)\}_{i=1}^N$. Therefore,
a BPS solition from $(0,v_j)$ to $(0,v_i)$ can be identified as a term of the
Floer differential of the pair of Lagrangian submanifolds, in the 
intersection Floer theory \cite{FOOO}.

\subsection{$k$ M2-branes}

We now consider the case of general $k$.
Let us take two ground states specified by subsets $V$ and $V'$ of
$\{v_i\}_{i=1}^N$ of order $k$.
A soliton that interpolates $V$ and $V'$ is the superposition of 
$k$ single M2-brane solitions, each of which approach
 $v_{i_a}\in V$ and $v_{i'_a}\in V'$ as $x^1\to -\infty$ and
$x^1\to +\infty$ respectively.
The central charge of such a solitonic sector is
the sum $\sum_{a=1}^kZ_{i'_a,i_a}$
while the mass is bounded below by
$\sum_{a=1}^k|Z_{i_a,i'_a}|$.
When $ u_j$ are generic, $Z_{ij}$ have different phases for
different pairs $(i,j)$.
Therefore, it saturates the BPS bound only when just one of the $k$ M2-branes
is a non-trivial soliton while the remaining $k-1$ stay fixed at the vacua.
This is possible only when $|V\cap V'|=k-1$.

In the picture where $\{v_i\}_{i=1}^N$ is regarded as the set of
weights of the fundamental representation $\bC^N$ of $\SU(N)$,
a BPS state for the $k$ M2-brane system
exists only if $V$ and $V'$, which are regarded as weights of
the representation $\wedge^k\bC^N$, are connected by
a root of $\SU(N)$. 
This matches with the structure of the BPS spectrum of the
Landau-Ginzburg model:
In \cite{Fendley:1990zj,Lerche:1991re}, it was proposed 
that there is exactly one BPS solition for each pair of vacua labelled by
weights of $\wedge^k\bC^N$ that differ by a root of $\SU(N)$. 
Therefore, we would like to  see that
there is exactly one BPS soliton for any pair of $v_i$ and $v_j$
in the single M2-brane system.

Showing this seems to be a difficult problem to the authors.
Instead of trying to find BPS configurations directly,
we shall take a certain limit  \cite{Mikhailov:1997jv}
that reduces the problem of finding BPS membranes to the
problem of finding BPS geodesics \cite{Klemm:1996bj},
and then use the technique of
spectral networks.

\subsection{BPS states via spectral networks}

\begin{figure}[htb]
	\centering
	\begin{subfigure}{.4\textwidth}
		\centering
		\includegraphics[width=\textwidth]{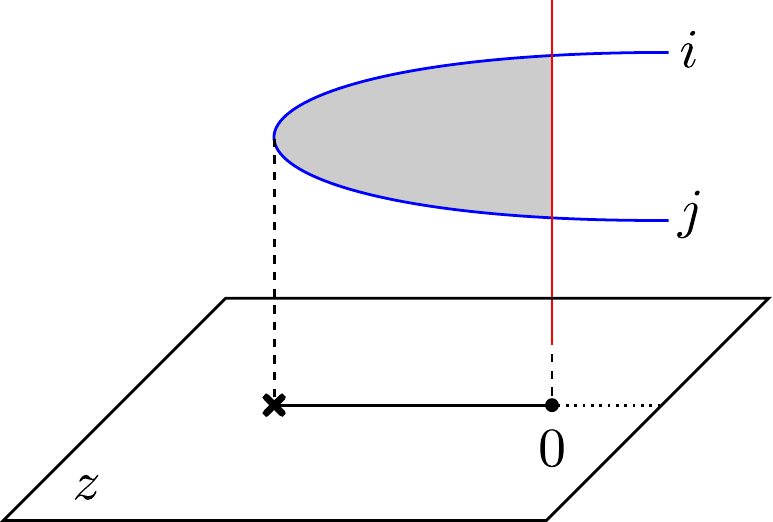}
		\caption{}
		\label{fig:M2soliton_A_1}
	\end{subfigure}
	\begin{subfigure}{.5\textwidth}
		\centering
		\includegraphics[width=\textwidth]{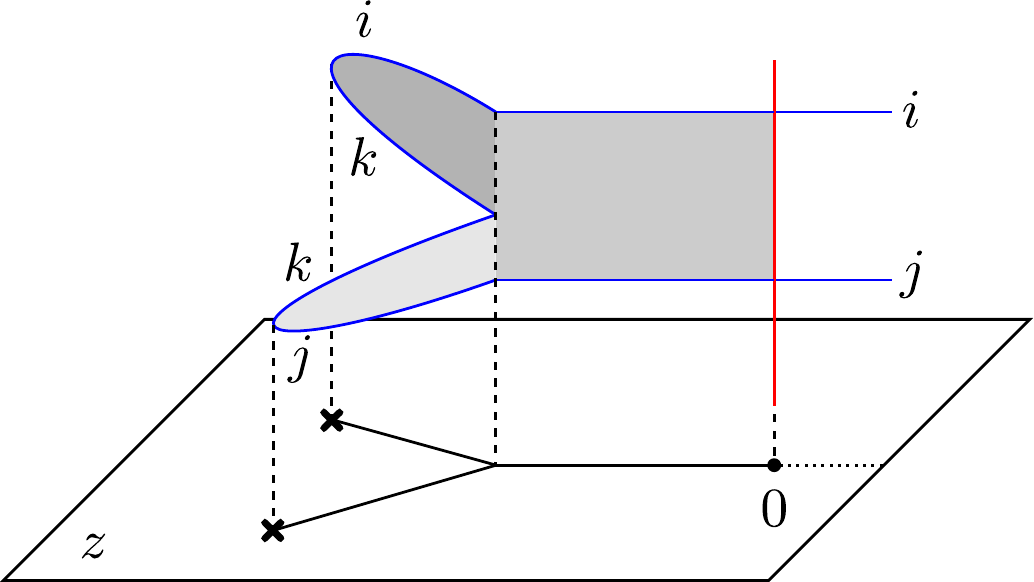}
		\caption{}
		\label{fig:M2soliton_A_2}
	\end{subfigure}
	\caption{M2-brane solitons in the $\beta\to 0$ limit. Blue curves are parts of M5, and red lines are parts of M5$'$. The $(z,v)$ images of the M2-brane solitons are shaded.}
	\label{fig:M2_soliton}
\end{figure}

So far, we have been using the metric 
$\dd s^2=|\dd z|^2+|\dd v|^2$ in the $x^{4,5,7,10}$ directions.
We now change it to
\beq
\dd s^2=|\dd z|^2+\beta^2|\dd v|^2
\eeq
and take a small $\beta$ limit. We also have
\beq
\omega={i\over 2}\dd z\wedge\dd\overline{z}
+{i\over 2}\beta^2\dd v\wedge\dd\overline{v},\quad\,\,
\Omega=\beta\dd z\wedge \dd v.
\eeq
The argument of \cite{Mikhailov:1997jv} shows that, in the limit $\beta\to 0$,
the projection of a BPS configuration $C_{ij}=C_{\phi_i,\phi_j}$
onto the $z$-plane is a real one-dimensional graph $\gamma_{ij}$,
and the tangent directions $\Delta z$ and $\Delta v$ obey the constraint
$\Delta z\cdot\Delta v=\zeta_{ij}\cdot$ (real number).
Over a generic point $z$ on the graph $\gamma_{ij}$, 
$C_{ij}$ is a line segment from one solution $v_l$ to another $v_k$
of (\ref{curve}).
Of course, $(k,l)=(i,j)$ near $z=0$, but that may not be the case if
$z$ is far from $z=0$. See Figure~\ref{fig:M2_soliton}.

In a neighborhood of such a point, the graph $\gamma_{ij}$ is a curve
determined by the differential equation
\begin{align}
	\lambda_{kl} (z) \frac{\partial z}{\partial \tau}
= \exp(i \vartheta_{ij}) = \frac{Z_{ij}}{|Z_{ij}|}, \label{eq:BPS M2 boundary}
\end{align}
where $\lambda_{kl}\dd z = (v_k(z) - v_l(z))\dd z$ is
the difference of $\lambda=v\dd z$ at the $k$-th sheet and at the $l$-th
sheet of M5-branes, and $\tau$ is a real parameter along the curve
$\gamma_{ij}$.
Such a $\gamma_{ij}$ is called a finite open web of BPS
strings \cite{Gaiotto:2012rg}.

We would like to find a solution to (\ref{eq:BPS M2 boundary})
which starts from a branch point of the covering $v(z)\mapsto z$ and call it
$\cS_{kl}$.
For a generic value of $\vartheta=\vartheta_{ij}$, 
it does not pass the endpoint of the ground-state M2-branes at
$x=0$ but goes to infinity,
meaning that it does not correspond to any of the BPS states.
These paths are called $\cS$-walls.  When two
$\mathcal{S}$-walls $\cS_{ik}$ and $\cS_{kj}$ cross, another
$\mathcal{S}$-wall, $\mathcal{S}_{ij}$, can emerge, when there is a
supersymmetric junction of three M2-branes that satisfy $\lambda_{ik}
+ \lambda_{kj} = \lambda_{ij}$
\cite{Gaiotto:2009hg,Gaiotto:2012rg},
like in Figure \ref{fig:M2soliton_A_2}. The collection of $\mathcal{S}$-walls is called a spectral network
\cite{Gaiotto:2012rg}.  When there is an $\cS$-wall $\cS_{ij}$ that
passes $z=0$, then this gives us a BPS object with a finite central
charge.

\subsubsection{Deformation by \texorpdfstring{$ u_N$}{u\_N}}\label{symmetric}

Let us consider a particular deformation where the curve is 
\begin{equation}
	 z =  v^N +  u_N.
\end{equation} 
The $z$-coordinate is zero when
\begin{align}
	v_j = (- u_N)^{1/N} \omega^j,\quad
	\text{where}\quad \omega=e^{2\pi i/N}.
\end{align} 
Hence, the vacua are depicted by the vertex of a regular polygon on the $v$-plane.
 
\begin{figure}[ht]
	\centering
	\includegraphics[scale=.5]{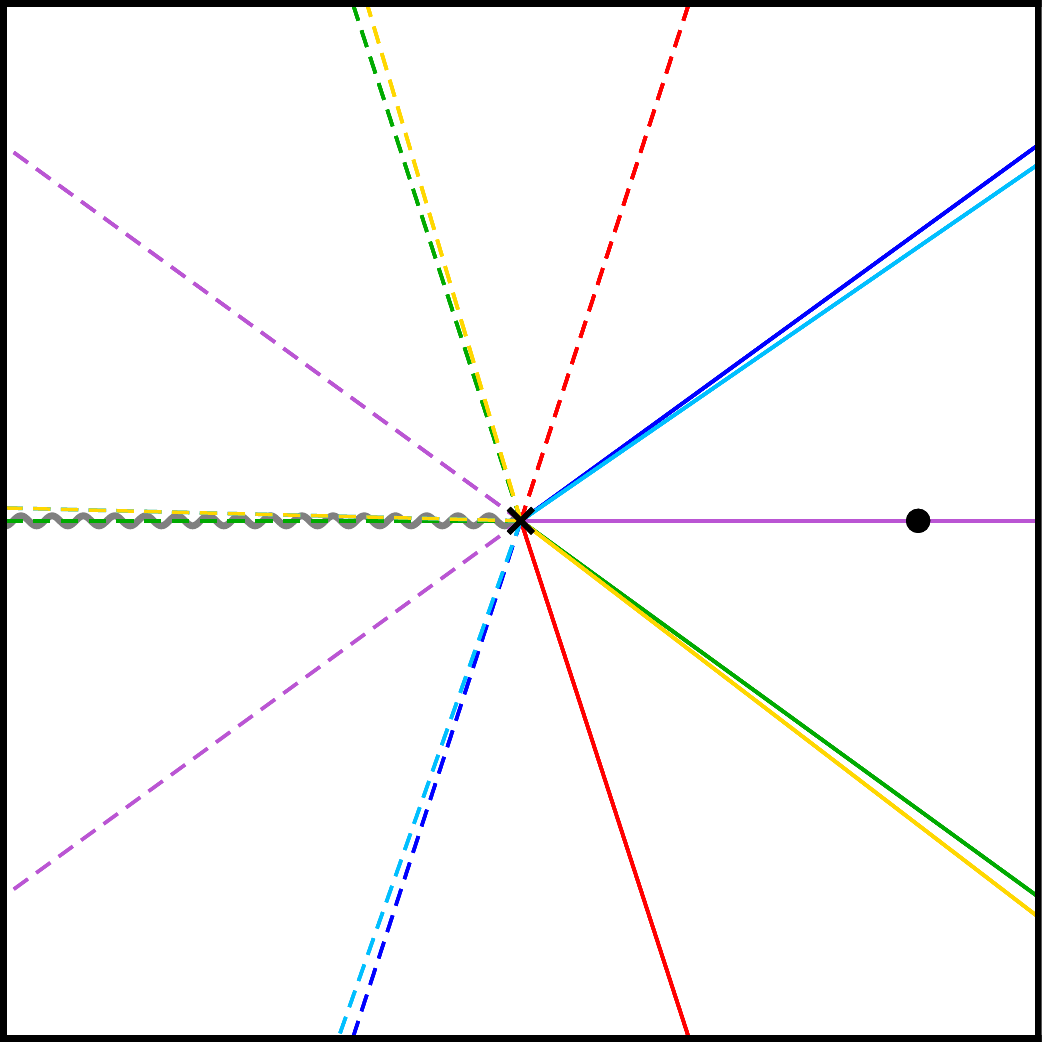}
	\caption{A spectral network around a branch point of ramification index $N=4$.}
	\label{fig:IRMN4k_S_Wall}
\end{figure}

 The curve has a branch point of ramification index $N$ at $z =
  u_N$, and the differential equation that governs the behavior of
 each $\mathcal{S}_{ij}$ on the $z$-plane is
\begin{align}
	\alpha_{ij} (z -  u_N)^{1/N} \frac{\partial z}{\partial \tau} = \exp(i \vartheta), \label{eq:differential equation for ij-string of M'(N,k,mu_N)}
\end{align}
where $\alpha_{ij} = \omega^i - \omega^j$.
The solution is
\begin{align}
	z_{ij}(\tau) =  u_N + c\left(\frac{N+1}{N} \frac{\tau}{\alpha_{ij}} \right)^{N/N+1} \exp \left( \frac{N}{N+1} i \vartheta \right) \label{eq:solution for an S_ij}
\end{align} 
where $c$ is an $(N+1)$-st root of unity.  This is a straight line
starting at the branch point $z= u_N$.  When
$\alpha_{ij}=\alpha_{i'j'}$ two S-walls can be on top of each other.
As an example, Figure \ref{fig:IRMN4k_S_Wall} shows the spectral network when $N=4$
for $\vartheta = 0$.  As can be seen there, $\cS_{12}$ and $\cS_{34}$
are coincident.

When we change $\vartheta$ from $0$ to $2\pi$, the whole spectral
network rotates by $2\pi N/(N+1)$, and the endpoint of the M2-brane
meets $N(N-1)$ $\mathcal{S}$-walls in the process, implying there are
in total $N(N-1)$ BPS states in the BPS spectrum of this
theory. Therefore, for each distinct $i$ and $j$, there is one BPS
state in the sector with the right boundary set to the vacuum $i$ and
the left boundary set to the vacuum $j$.  It is easy to identify the
value $\vartheta$ when an $\cS_{ij}$ wall hits $x=0$. There is one
value of $\vartheta$ for each $\cS_{ij}$.

\begin{figure}[ht]
	\centering
	\begin{subfigure}{.4\textwidth}	
	\centering
		\vspace{23pt}
		\includegraphics[width=\textwidth]{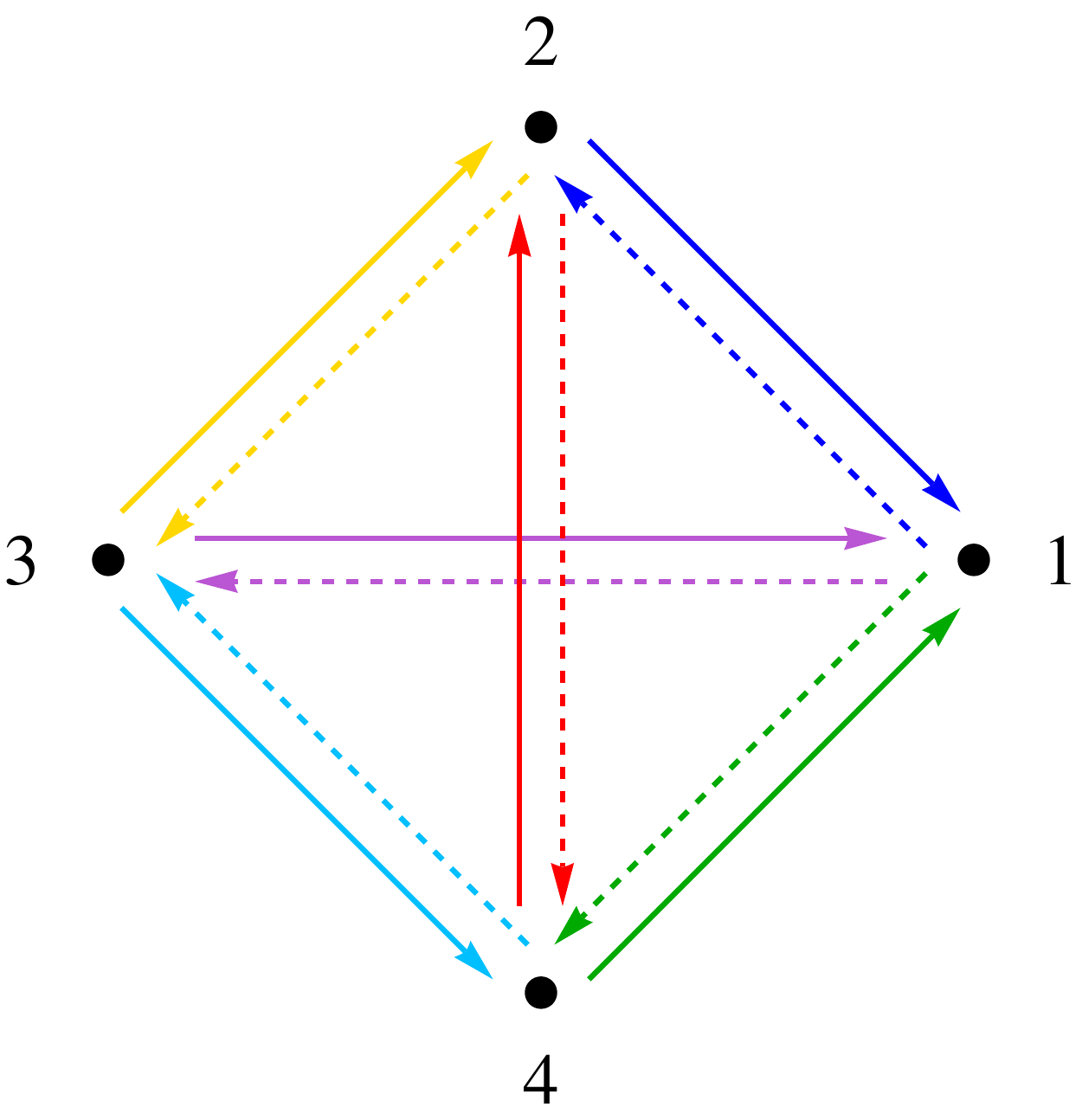}
		\vspace{10pt}
		\caption{on the $v$-plane}
		\label{fig:IRMN4k1_diagram_v_plane}
	\end{subfigure}
	\hspace{.1\textwidth}
	\begin{subfigure}{.4\textwidth}	
		\centering
		\includegraphics[width=\textwidth]{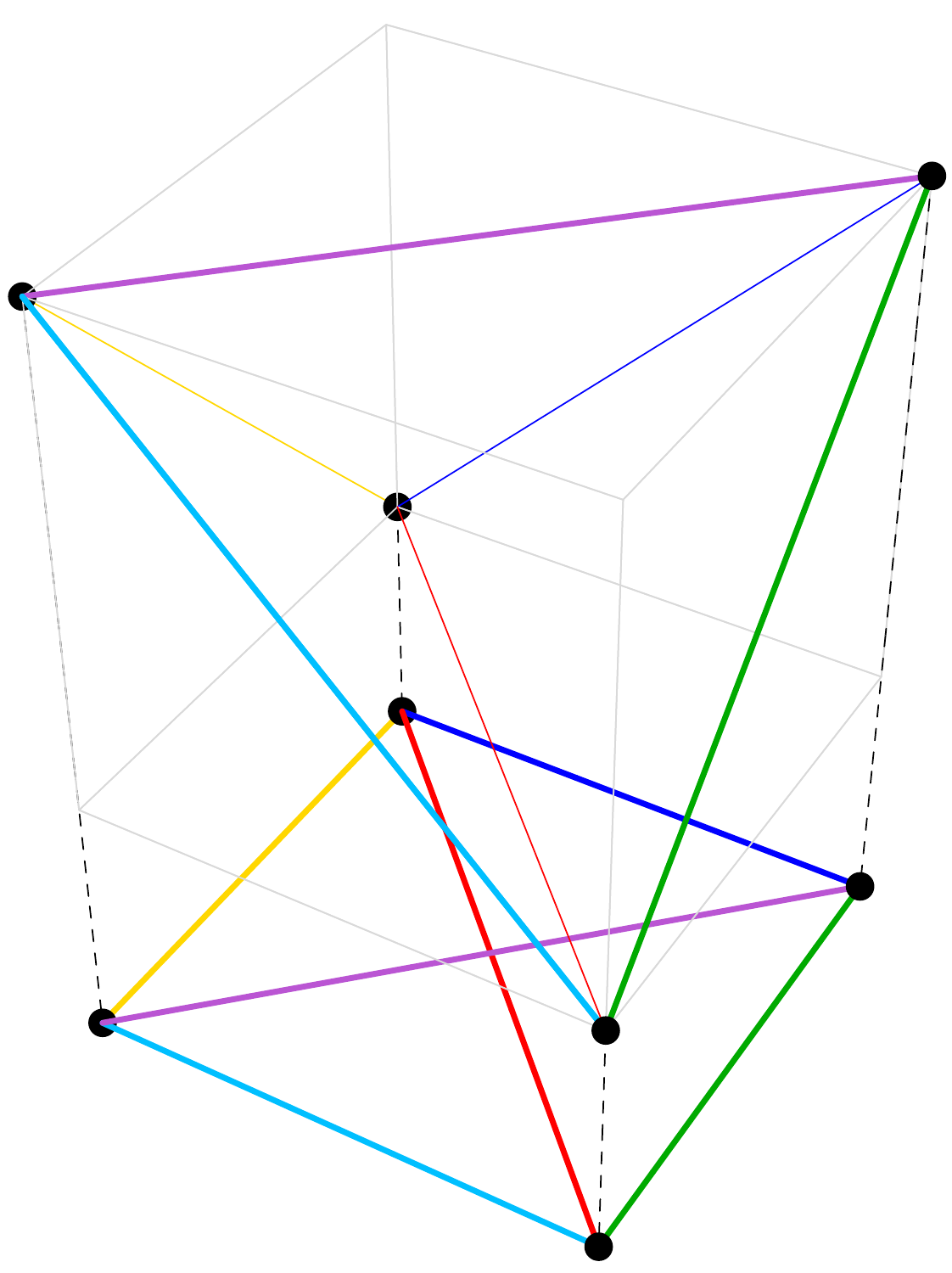}
		\caption{in the weight lattice of $\SU(4)$}
		\label{fig:IRMN4k1_diagram_weight_lattice}
	\end{subfigure}
	\caption{Vacua and solitons, $k=1$.}
	\label{fig:IRMN4k1_diagram}
\end{figure}

On the $v$-plane, we can introduce a soliton of the $k=1$ theory by a
line connecting $v_i$ and $v_j$. Let us illustrate the case $N=4$.
Figure \ref{fig:IRMN4k1_diagram_v_plane} represents the four ground
states and twelve solitons on the $v$-plane.  We clearly see that
$Z[\gamma_{12}]$ and $Z[\gamma_{34}]$ has the same phase, as was also
reflected in the spectral network shown in
Figure~\ref{fig:IRMN4k_S_Wall}. Note that Figure
\ref{fig:IRMN4k1_diagram_v_plane} can be understood as obtained from
the projection of the weights of the fundamental representation of
$\SU(4)$ and the roots connecting the weights, representing the ground
states and the solitons respectively, as shown in Figure
\ref{fig:IRMN4k1_diagram_weight_lattice}. This structure of BPS
solitons is the same as that of the corresponding Landau-Ginzburg
model with a single chiral field, which has as its IR fixed point the
$\cN{=}2$ $A_3$ minimal model \cite{Fendley:1990zj}.

\begin{figure}[ht]
	\centering
	\begin{subfigure}{.45\textwidth}	
		\includegraphics[width=\textwidth]{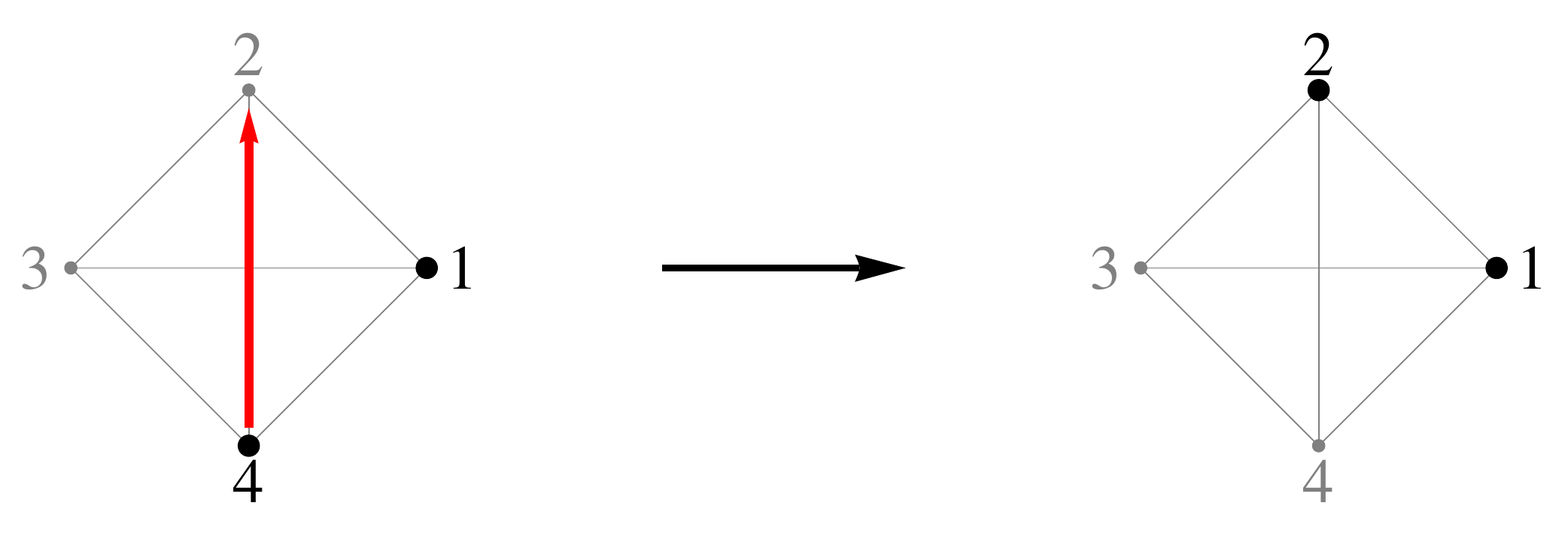}
		\caption{$[14] \to [12]$ soliton for $k=2$}
		\label{fig:IRMN4k2_soliton_diagram_01}
	\end{subfigure}
	\hspace{10pt}
	\begin{subfigure}{.45\textwidth}	
		\includegraphics[width=\textwidth]{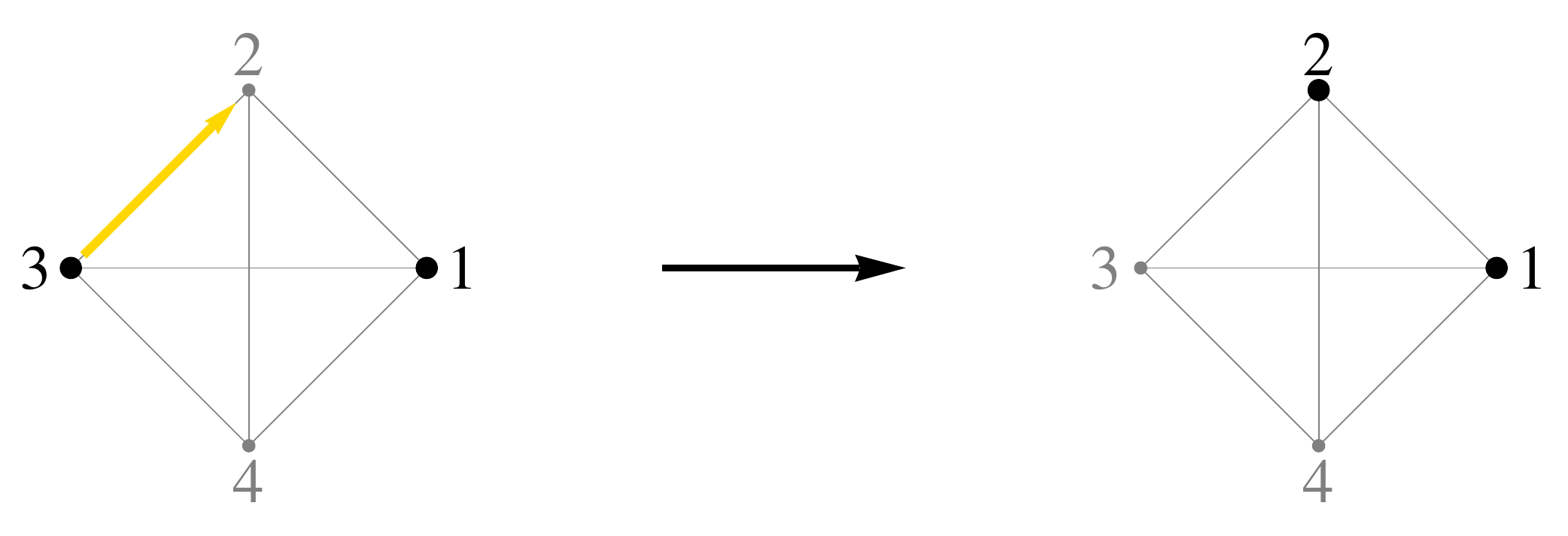}
		\caption{$[13] \to [12]$ soliton for $k=2$}
		\label{fig:IRMN4k2_soliton_diagram_02}
	\end{subfigure}
	\begin{subfigure}{.45\textwidth}	
	\includegraphics[width=\textwidth]{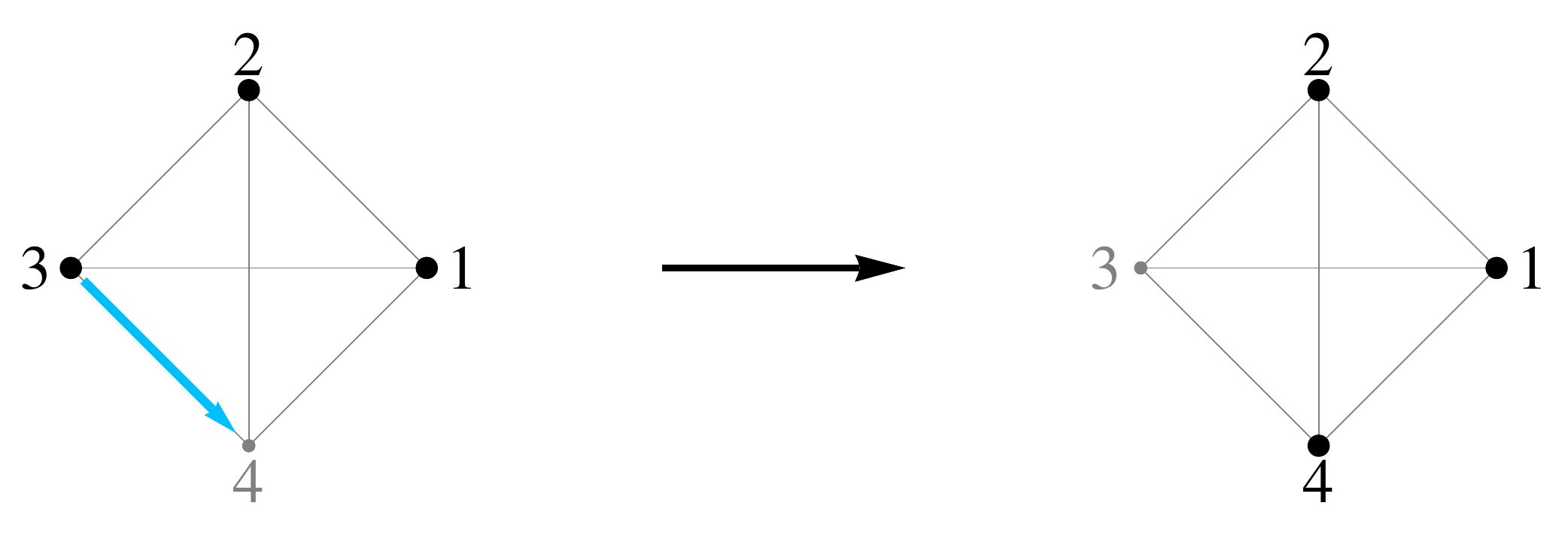}
	\caption{$[123] \to [412]$ soliton for $k=3$}
	\label{fig:IRMN4k3_soliton_diagram}
	\end{subfigure}
	\caption{Examples of $k > 1$ solitons.}
	\label{fig:IRMN4k}
\end{figure}

\begin{figure}[ht]
	\centering
	\begin{subfigure}{.4\textwidth}	
	\centering
		\vspace{20pt}
		\includegraphics[width=\textwidth]{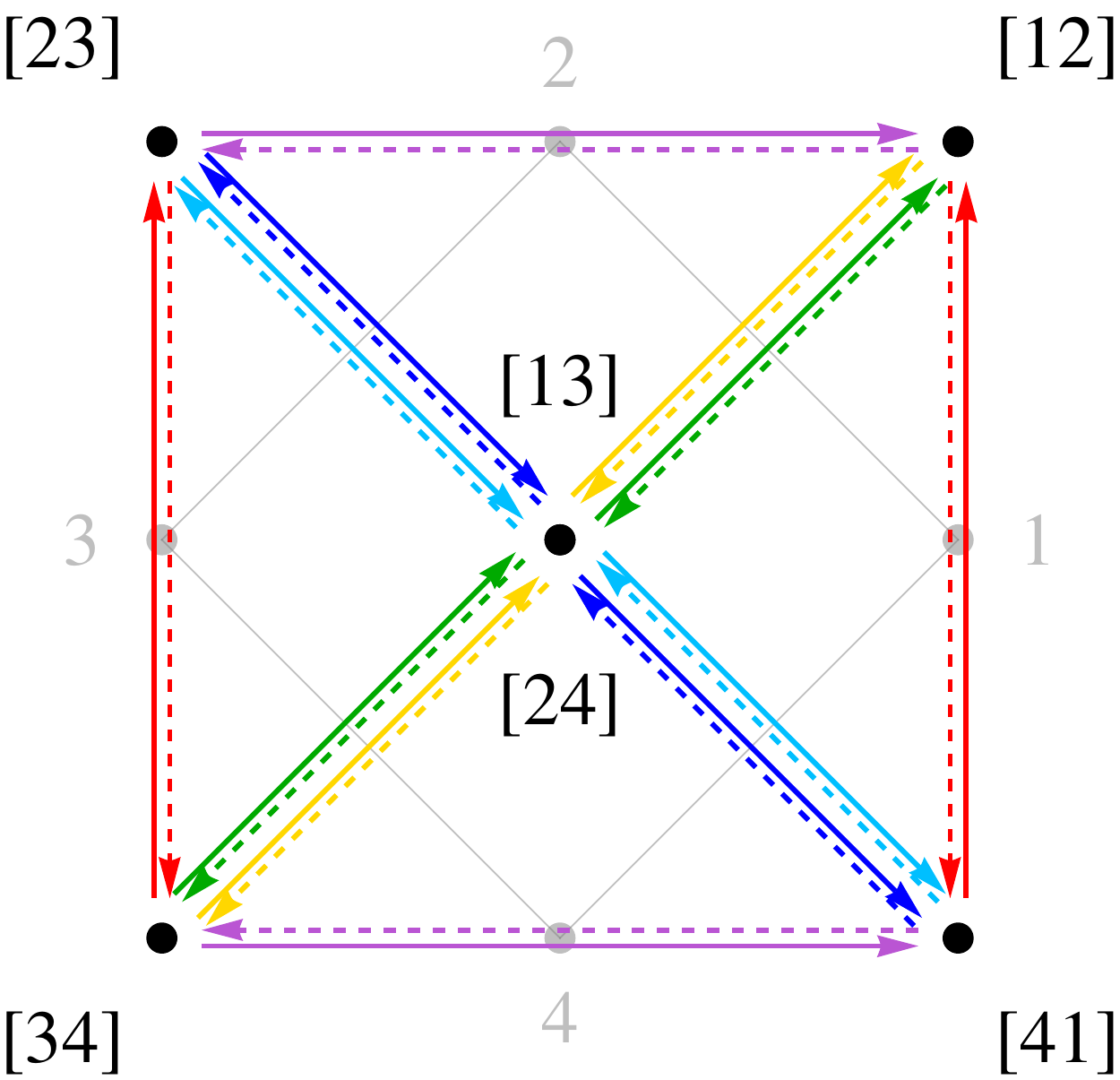}
		\vspace{4pt}
		\caption{on the $v$-plane}
		\label{fig:IRMN4k2_diagram_v_plane}
	\end{subfigure}
	\hspace{.1\textwidth}
	\begin{subfigure}{.4\textwidth}	
		\centering
		\includegraphics[width=\textwidth]{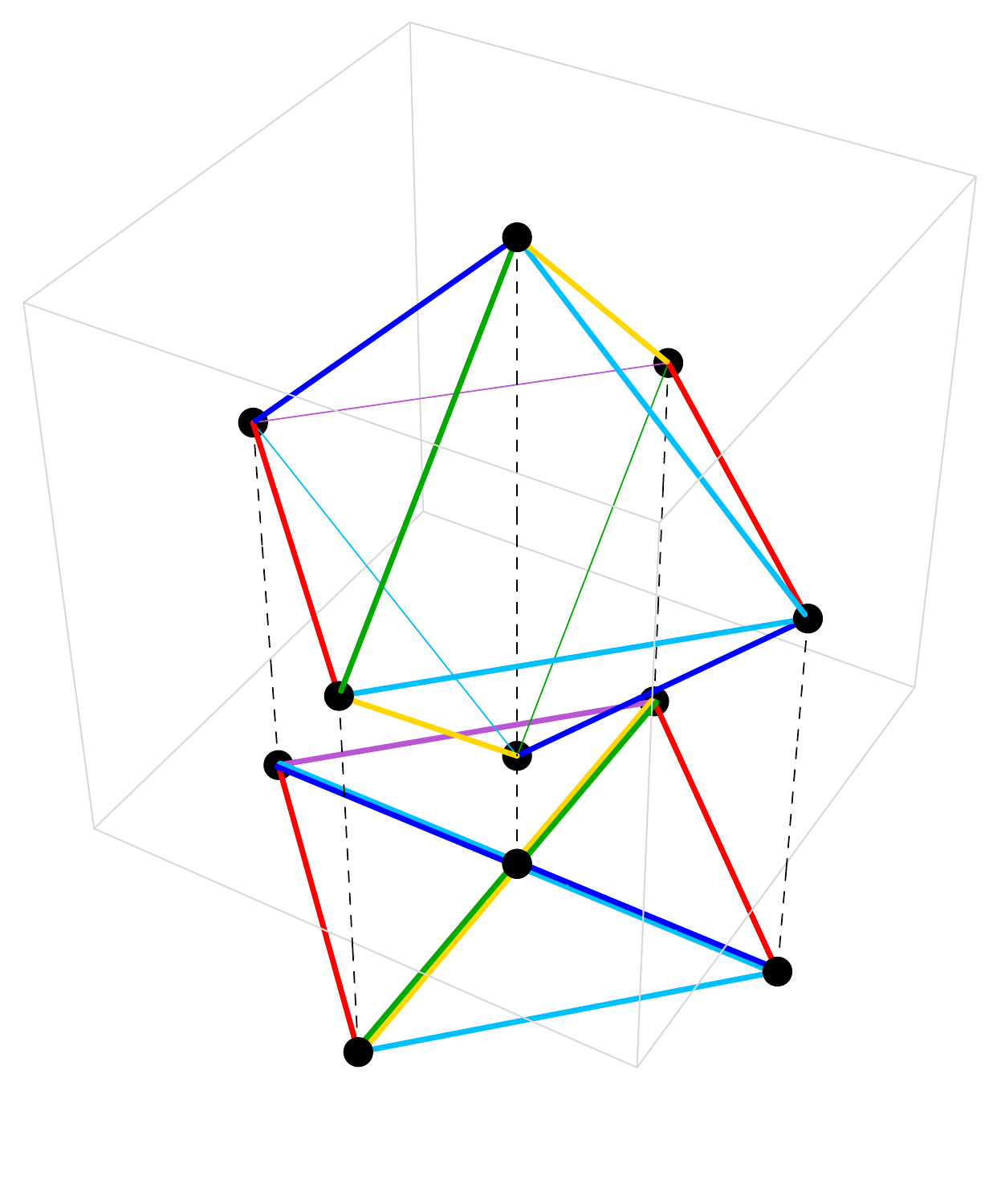}
		\caption{in the weight lattice of $\SU(4)$}
		\label{fig:IRMN4k2_diagram_weight_lattice}
	\end{subfigure}
	\caption{Vacua and solitons, $k=2$.}
	\label{fig:IRMN4k2_diagram}
\end{figure}

So far we discussed the case when there is just one M2-brane,
$k=1$. For general $k$, we need to choose $k$ vertices out of $N$, and
a soliton is obtained by moving one of the $k$ vertices. In Figure
\ref{fig:IRMN4k}, some representative examples of the solitons with
$k=2$ and $k=3$ are shown.
For $k=2$, we see from Figures \ref{fig:IRMN4k2_soliton_diagram_01}
and \ref{fig:IRMN4k2_soliton_diagram_02} that a $k=1$ solitonic
configuration can connect two $k=2$ ground states. From this
consideration we can represent $k=2$ ground states and solitons as
shown in Figure \ref{fig:IRMN4k2_diagram_v_plane}. Again, we can
understand this as obtained from the projection of the weights of the
2nd antisymmetric power of the fundamental representation of $\SU(4)$
and the roots connecting the weights, as shown in Figure
\ref{fig:IRMN4k1_diagram_weight_lattice}. The same structure of BPS
solitons of the corresponding Landau-Ginzburg model is observed in
\cite{Lerche:1991re}, which is expected to flow in the IR to the
Kazama-Suzuki model based on $\SU(4)_1/{\mathrm{S}[\UU(2)\times
  \UU(2)]}$.

For $k=3$, because choosing $k$ ground states among $N$
indistinguishable ones is the same as choosing $N-k$ ground state, the
ground states and the solitons are represented by the same diagram as
Figure \ref{fig:IRMN4k1_diagram_v_plane}, thus we see the $k
\leftrightarrow N-k$ duality.

\subsubsection{General deformations}
Now let us consider how the spectral networks look when the
deformation parameters $ u_j$ are general.

\subsubsection*{BPS spectrum with $z=v^3$ }

\begin{figure}[ht]
	\centering
		\includegraphics[width=.3\textwidth]{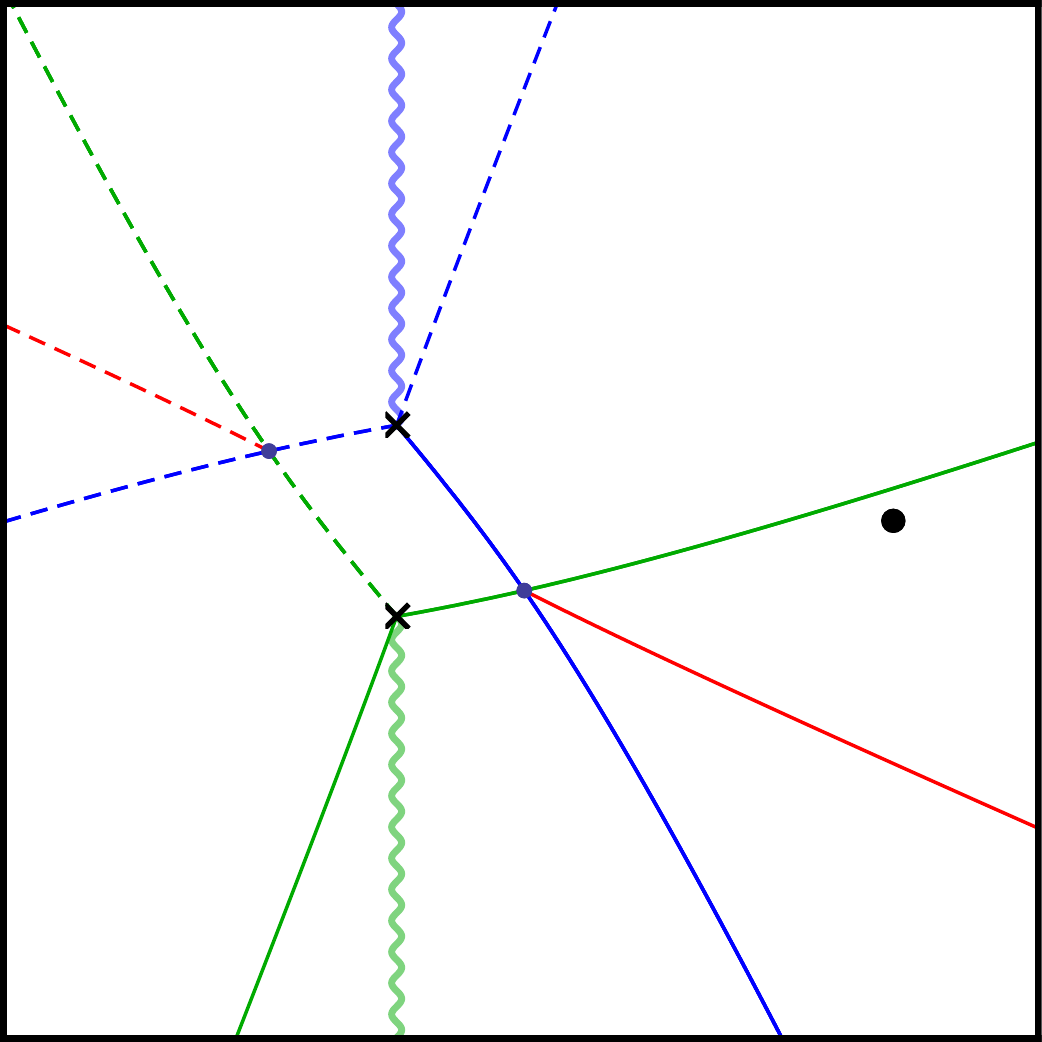}
		\caption{A spectral network with general $u_2$ and $u_3$.}
		\label{fig:IRMN3k_SWall_after_joint}
\end{figure}

Now we consider the case where we have three M5-branes ramified over
the $z$-plane:
\begin{align}
	 z =   v^3 +  u_2 v +  u_3.
\end{align}
For general $ u_2$ and $ u_3$, we have two branch points of
ramification index 2 on the $z$-plane as shown in Figure
\ref{fig:IRMN3k_SWall_after_joint}. In the figure, we chose
$ u_{2,3}$ so that a (12)-branch cut, a blue wavy line, comes out
from the upper branch point, and (13)-branch cut, a green wavy line,
from the lower branch point.  From the (12)-branch point we have three
$\mathcal{S}$-walls: two $\mathcal{S}_{21}$ with solid blue line and
one $\mathcal{S}_{12}$ with a dashed blue line.  Similarly, from the
(13)-branch point, we have two $\mathcal{S}_{13}$ with solid green
line and one $\mathcal{S}_{31}$ with dashed green line.  We can see
that one $\mathcal{S}_{21}$ and one $\mathcal{S}_{13}$ meet at a
point, from which another $\mathcal{S}$-wall, $\mathcal{S}_{23}$,
emerges.

\begin{figure}[ht]
	\centering
	\begin{subfigure}{.22\textwidth}
		\centering
		\includegraphics[width=\textwidth]{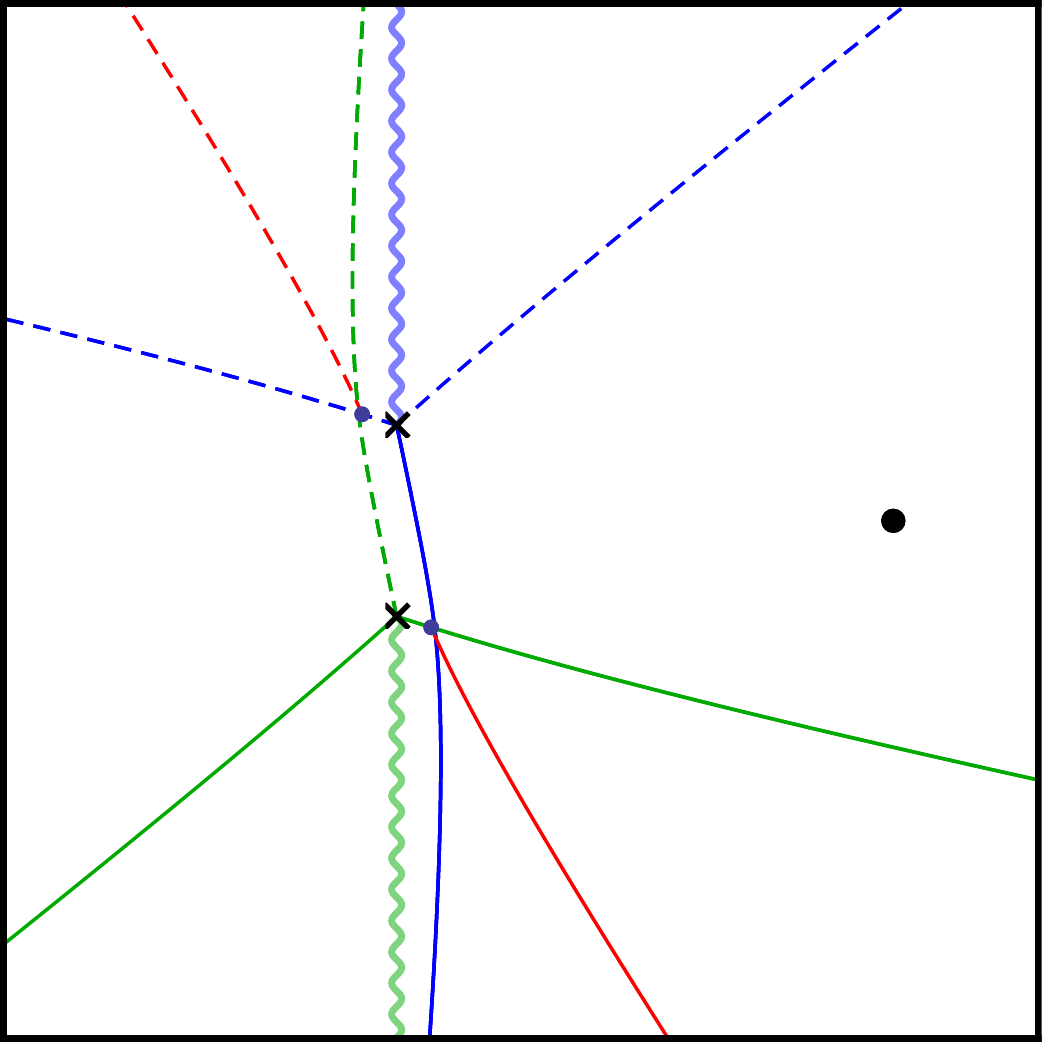}
		\caption{$\vartheta \approx 0$}
		\label{fig:IRMN3k_SWall_0}
	\end{subfigure}
\hspace{5pt}	
	\begin{subfigure}{.22\textwidth}
		\centering
		\includegraphics[width=\textwidth]{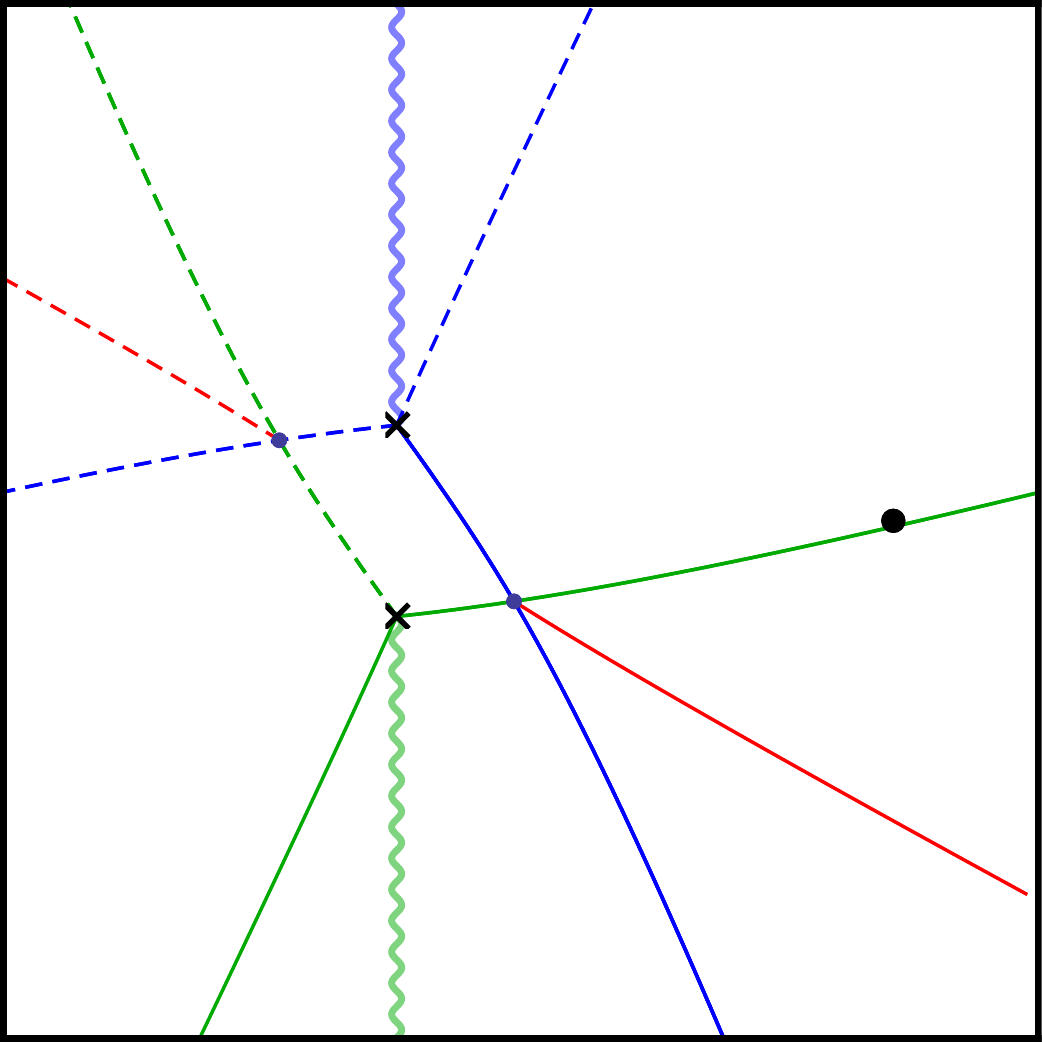}
		\caption{$\vartheta \approx \vartheta_{13}$}
		\label{fig:IRMN3k_SWall_13}
	\end{subfigure}
\hspace{5pt}	
	\begin{subfigure}{.22\textwidth}
		\centering
		\includegraphics[width=\textwidth]{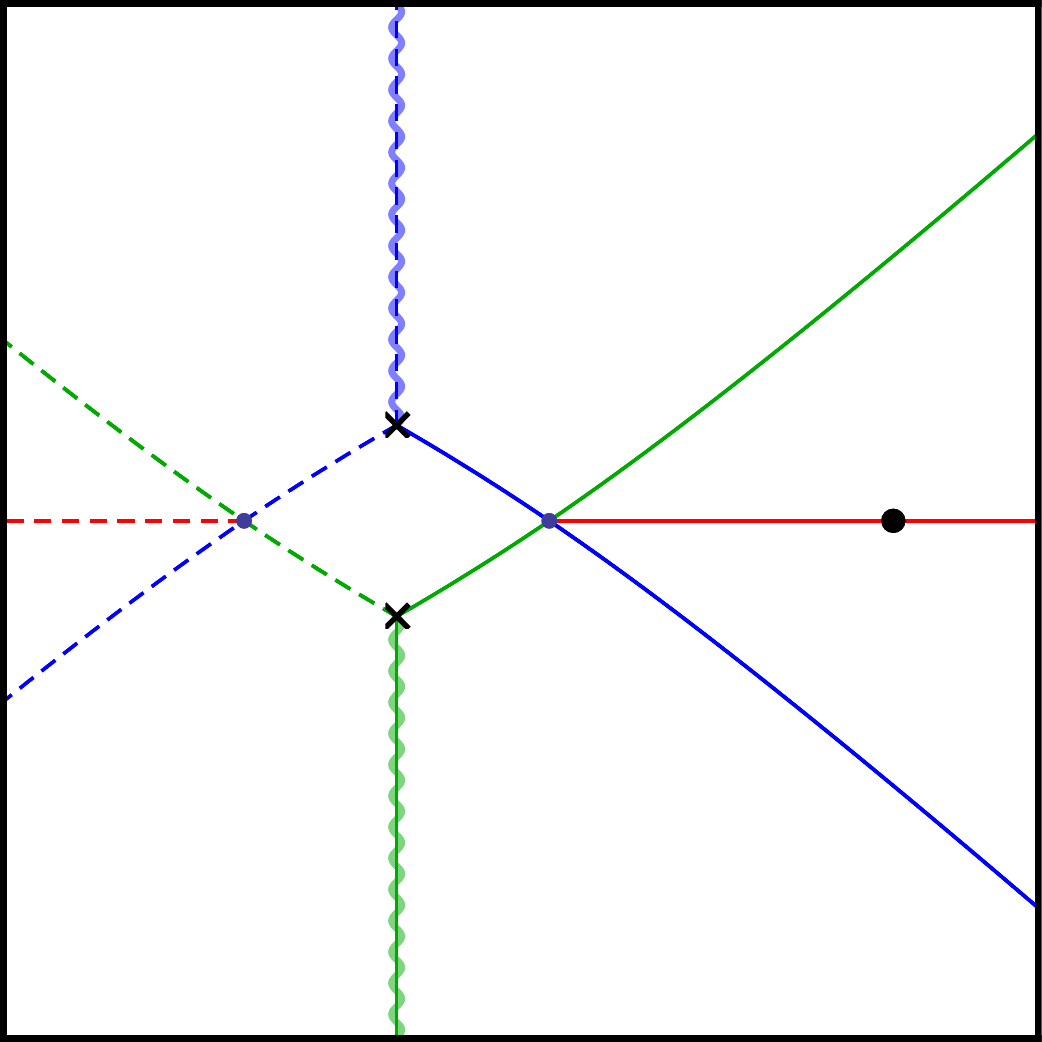}
		\caption{$\vartheta \approx \vartheta_{23}$}
		\label{fig:IRMN3k_SWall_23}
	\end{subfigure}
\hspace{5pt}
	\begin{subfigure}{.22\textwidth}
		\centering
		\includegraphics[width=\textwidth]{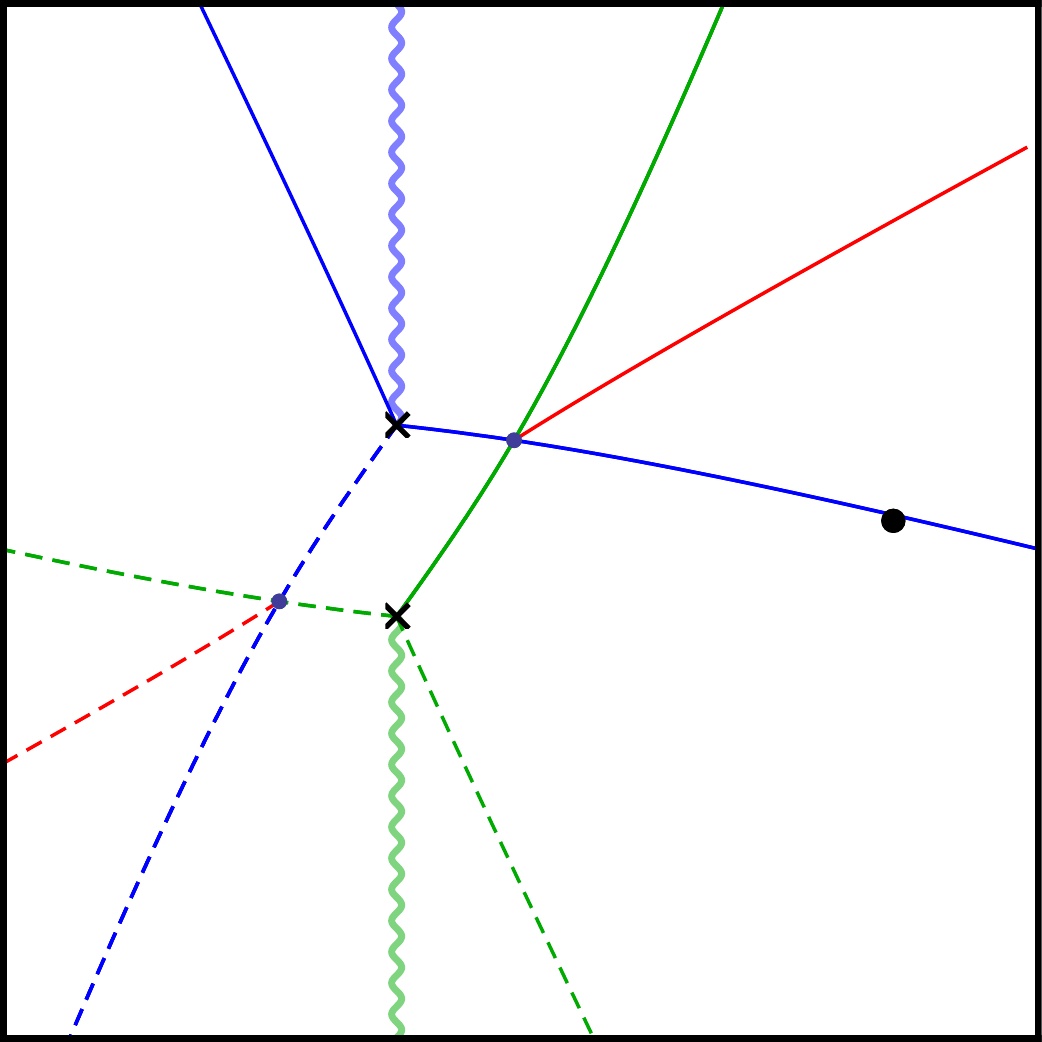}
		\caption{$\vartheta \approx \vartheta_{21}$}
		\label{fig:IRMN3k_SWall_21}
	\end{subfigure}
	
	\caption{Rotation of a spectral network with general $u_2$ and $u_3$.}
	\label{fig:IRMN3k_SWall}
\end{figure}

As we now have the full spectral network, let us rotate it by changing
$\vartheta$ from $0$ to $2\pi$.  Figure \ref{fig:IRMN3k_SWall} shows
spectral networks at various values of $\vartheta$, $0 <
\vartheta_{13} < \vartheta_{23} < \vartheta_{21} < \pi$. We see that
there are $\gamma_{13}$, $\gamma_{23}$, and $\gamma_{21}$ at
$\vartheta_{13}$, $\vartheta_{23}$, and $\vartheta_{21}$,
respectively, between the branch point and the M2-brane
endpoint. Therefore there are corresponding three BPS states for $0 <
\vartheta < \pi$. There are another three BPS states for $\pi <
\vartheta < 2\pi$, each of which has the central charge
$Z[\gamma_{ji}] = - Z[\gamma_{ij}]$.

\begin{figure}[ht]
	\centering
	\begin{subfigure}{.3\textwidth}
		\centering
		\includegraphics[width=\textwidth]{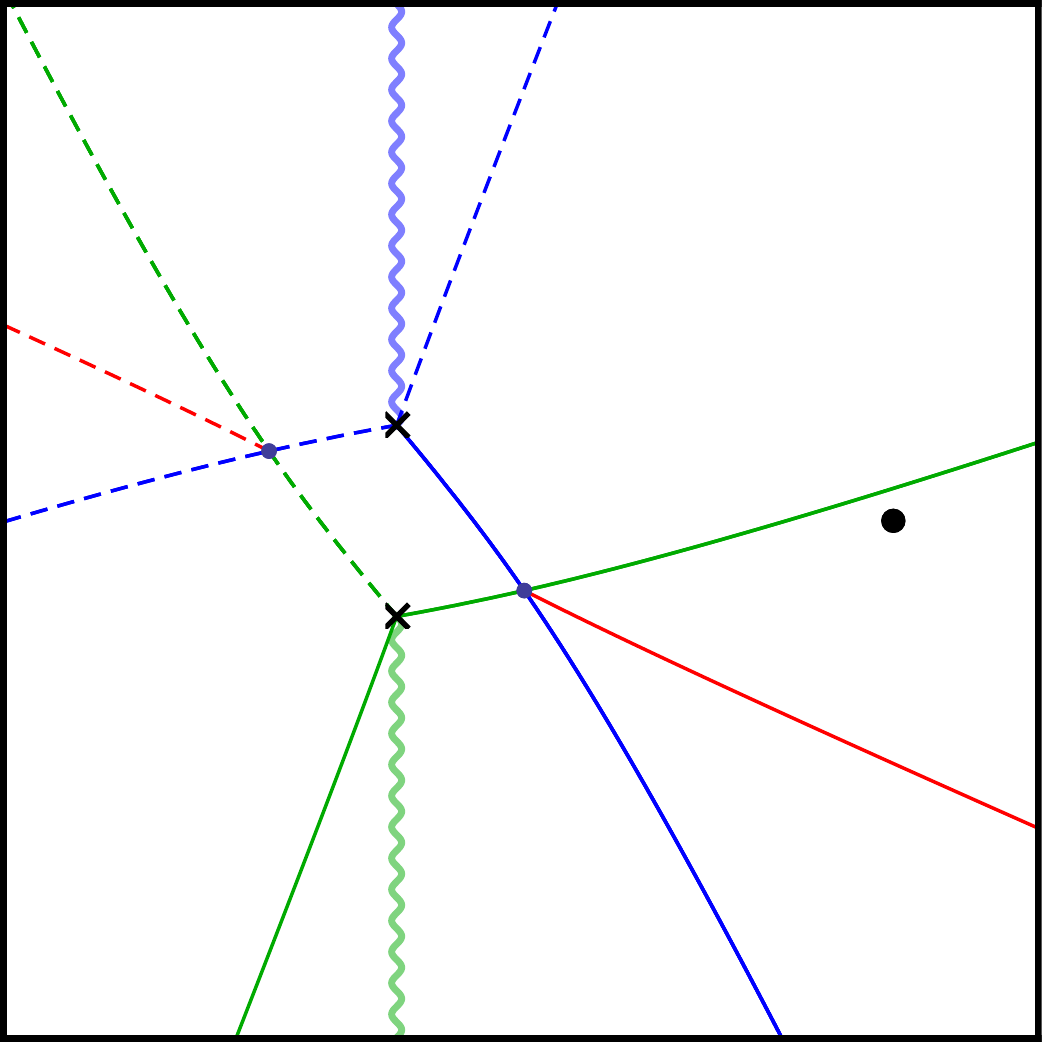}
		\caption{$ u_2 \approx 1$}
		\label{fig:IRMN3k_SWall_mu2_large}
	\end{subfigure}
\hspace{5pt}
	\begin{subfigure}{.3\textwidth}
		\centering
		\includegraphics[width=\textwidth]{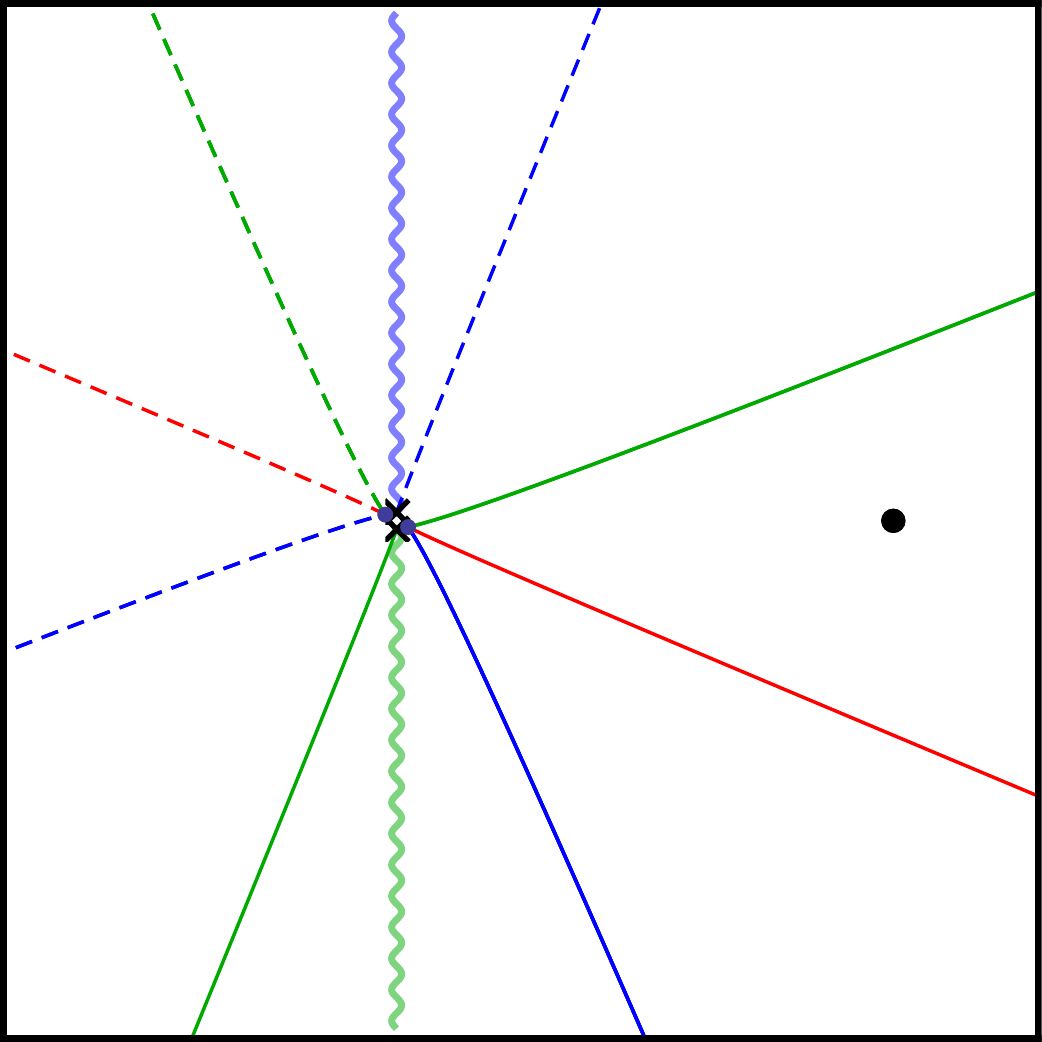}
		\caption{$ u_2 \ll 1$}
		\label{fig:IRMN3k_SWall_mu2_small}
	\end{subfigure}
\hspace{5pt}
	\begin{subfigure}{.3\textwidth}
		\centering
		\includegraphics[width=\textwidth]{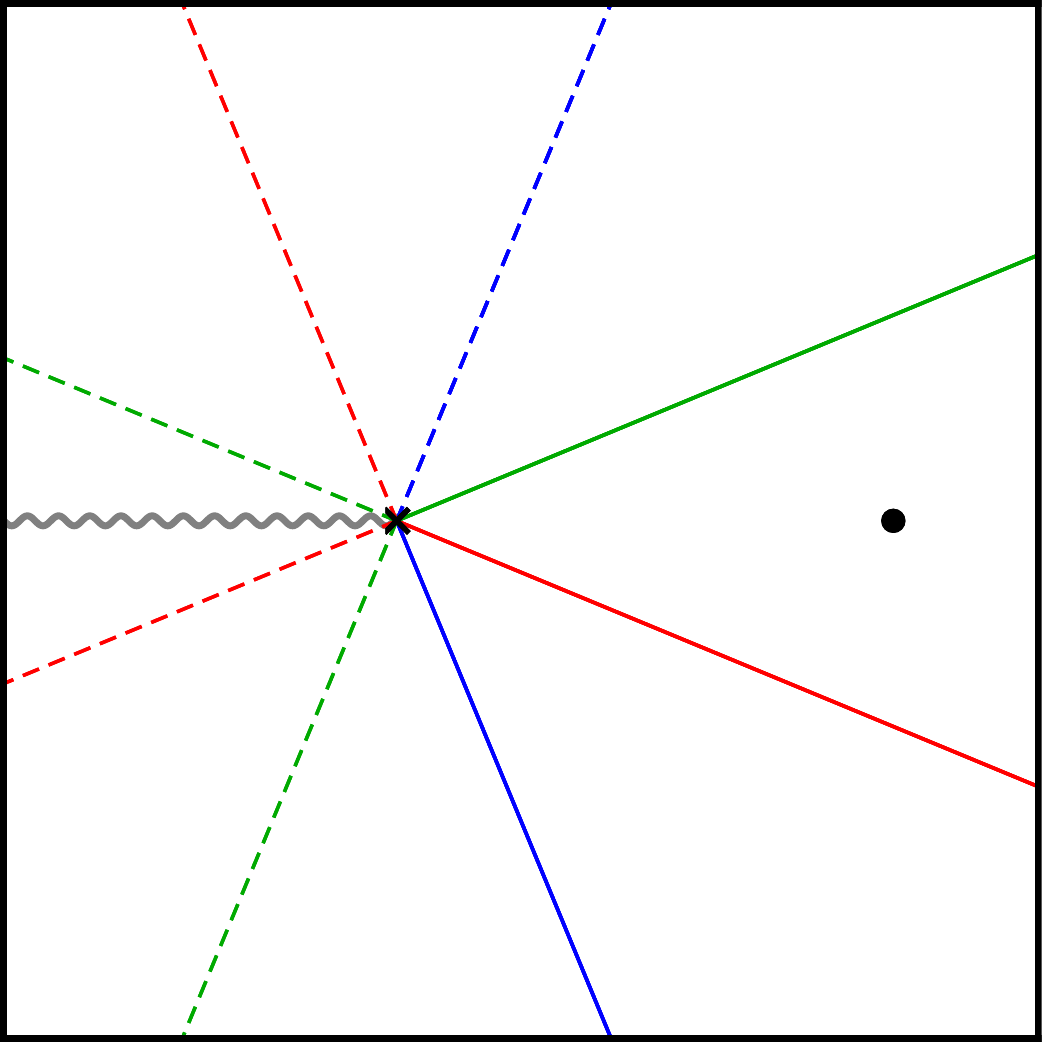}
		\caption{$ u_2 = 0$}
		\label{fig:IRMN3k_SWall_mu2_zero}
	\end{subfigure}
	
	\caption{Evolution of the spectral network under the limit of $u_2 \to 0$.}
	\label{fig:IRMN3k_SWall_mu2_vanishing}
\end{figure}

Now let us take the limit $ u_2 \to 0$ so that the two branch points
collide, see Figure \ref{fig:IRMN3k_SWall_mu2_vanishing}.
Figure \ref{fig:IRMN3k_SWall_mu2_zero} shows the spectral network when
$ u_2=0$.  There is only a single (123)-branch point and a single
(123)-branch cut.\footnote{The notation is that around the
  $(n_1n_2\cdots n_k)$ branch cut the sheets are exchanged in the
  order $n_1\to n_2\to \cdots \to n_k \to n_1$.}  The whole spectral
network rotates by $3\pi/4$ when we change $\vartheta$ from $0$ to
$\pi$ continuously, and in the process we find three BPS strings
connecting the branch point and the endpoint of the M2-brane,
corresponding to three BPS states in $0 < \arg(Z) < \pi$.

\subsubsection*{BPS spectrum with $z=v^4$}

Let us now consider the case $N=4$, for more illustration.

\begin{figure}[ht]
	\centering
	\begin{subfigure}{.4\textwidth}
		\centering
		\includegraphics[width=\textwidth]{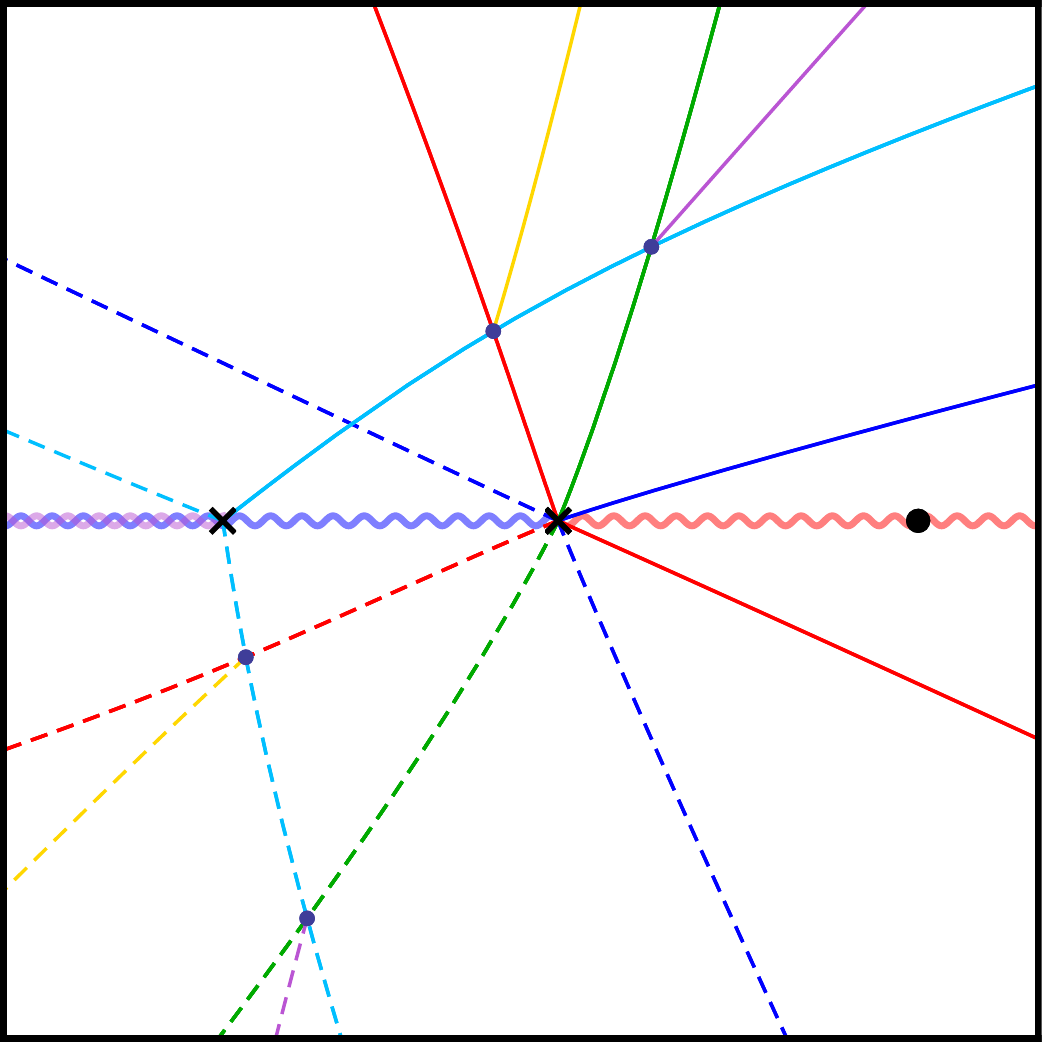}
		\label{fig:IRMN4k_SWall_after_joint}
	\end{subfigure}
	\begin{subfigure}{.1\textwidth}
		\includegraphics[width=.75\textwidth]{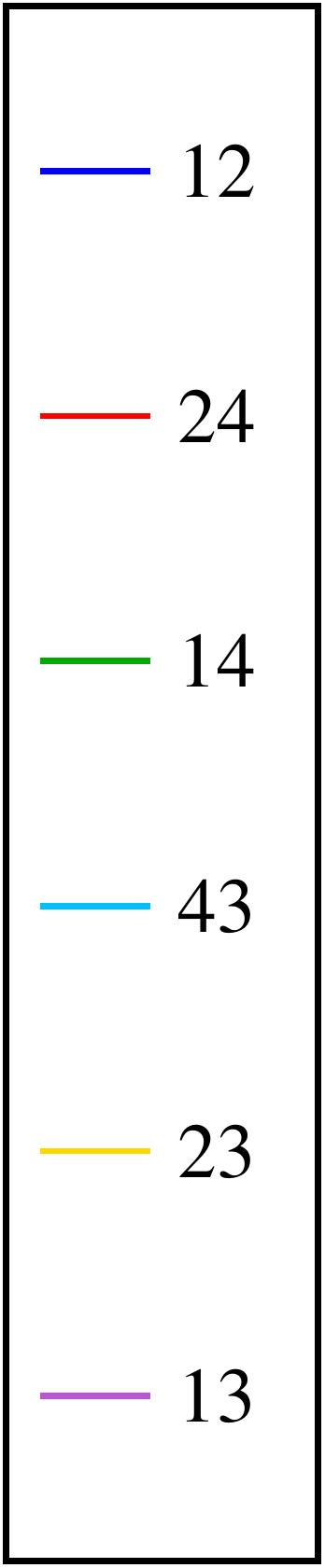}
		\vspace{20pt}
	\end{subfigure}
	\caption{A spectral network with $u_{2,3,4} \neq 0$.}
	\label{fig:IRMN4k_SWall_joint}
\end{figure}

Figure \ref{fig:IRMN4k_SWall_joint} shows the spectral network with
$ u_{2,3,4}$ chosen so that there is a $(124)$-branch point and a
$(34)$-brach point. See the legend for the nature of walls represented
by the colors and the styles.

\begin{figure}[ht]
	\centering
	\begin{subfigure}{.3\textwidth}
		\centering
		\includegraphics[width=\textwidth]{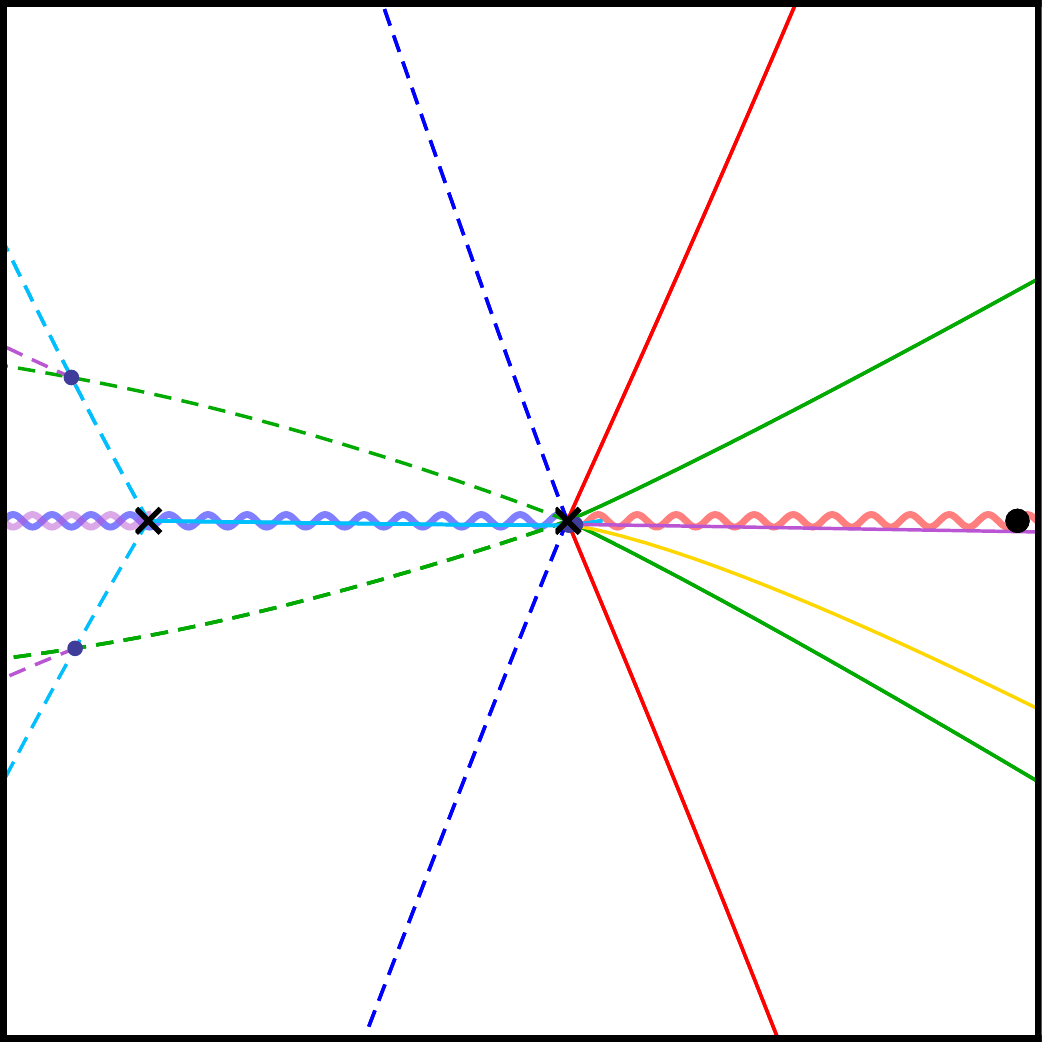}
		\caption{$\vartheta \approx \vartheta_{13}$}
		\label{fig:IRMN4k_SWall_01}
	\end{subfigure}
	\begin{subfigure}{.3\textwidth}
		\centering
		\includegraphics[width=\textwidth]{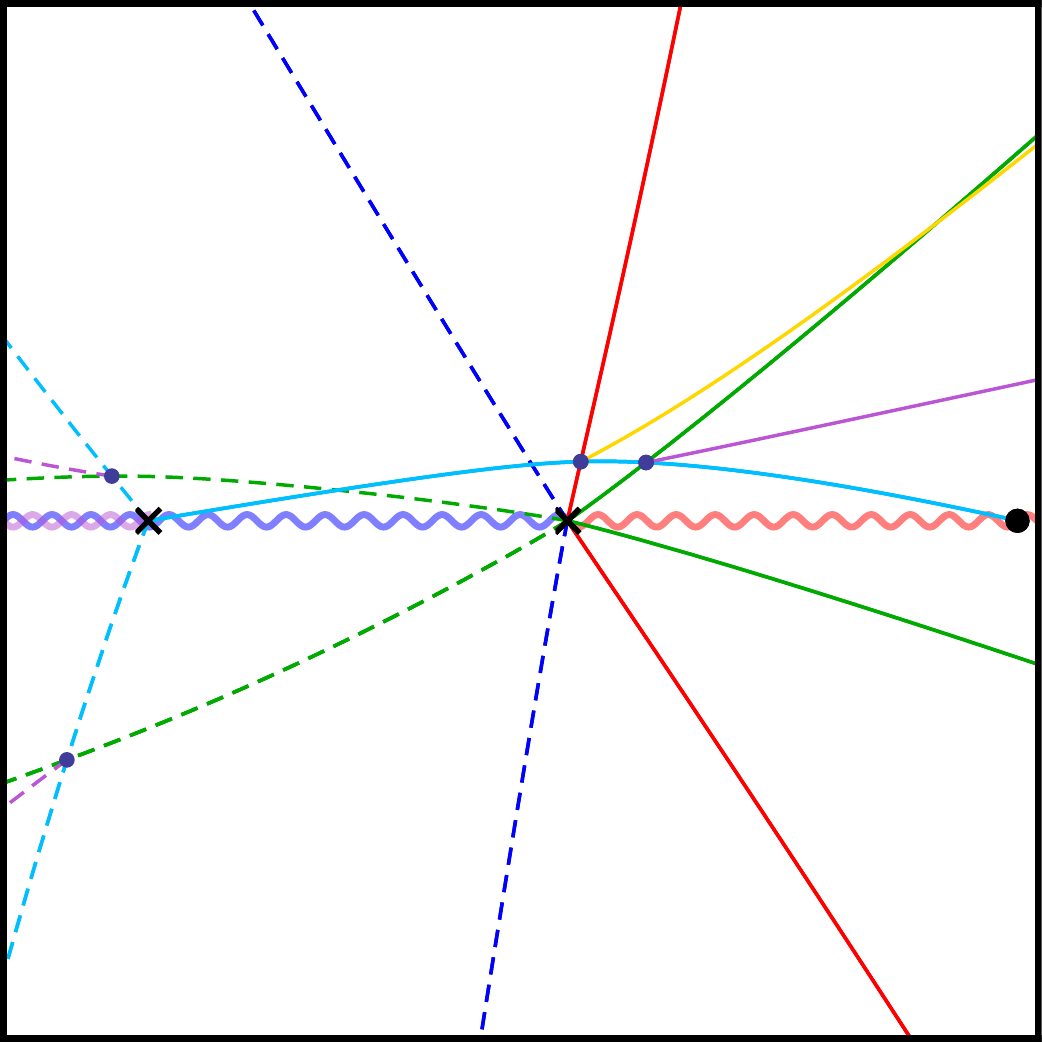}
		\caption{$\vartheta \approx \vartheta_{43}$}
		\label{fig:IRMN4k_SWall_02}
	\end{subfigure}
	\begin{subfigure}{.3\textwidth}
		\centering
		\includegraphics[width=\textwidth]{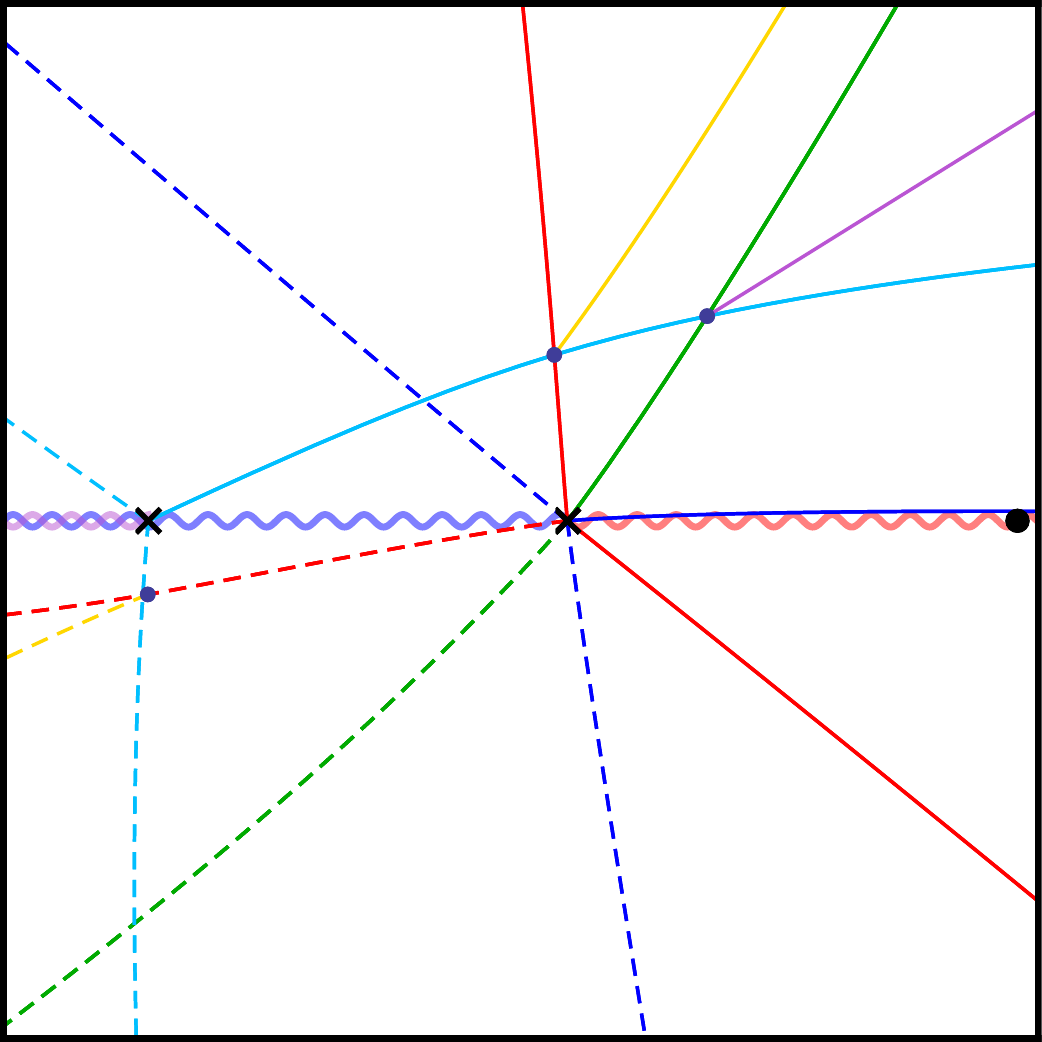}
		\caption{$\vartheta \approx \vartheta_{12}$}
		\label{fig:IRMN4k_SWall_03}
	\end{subfigure}
	
\vspace{10pt}	

	\begin{subfigure}{.3\textwidth}
		\centering
		\includegraphics[width=\textwidth]{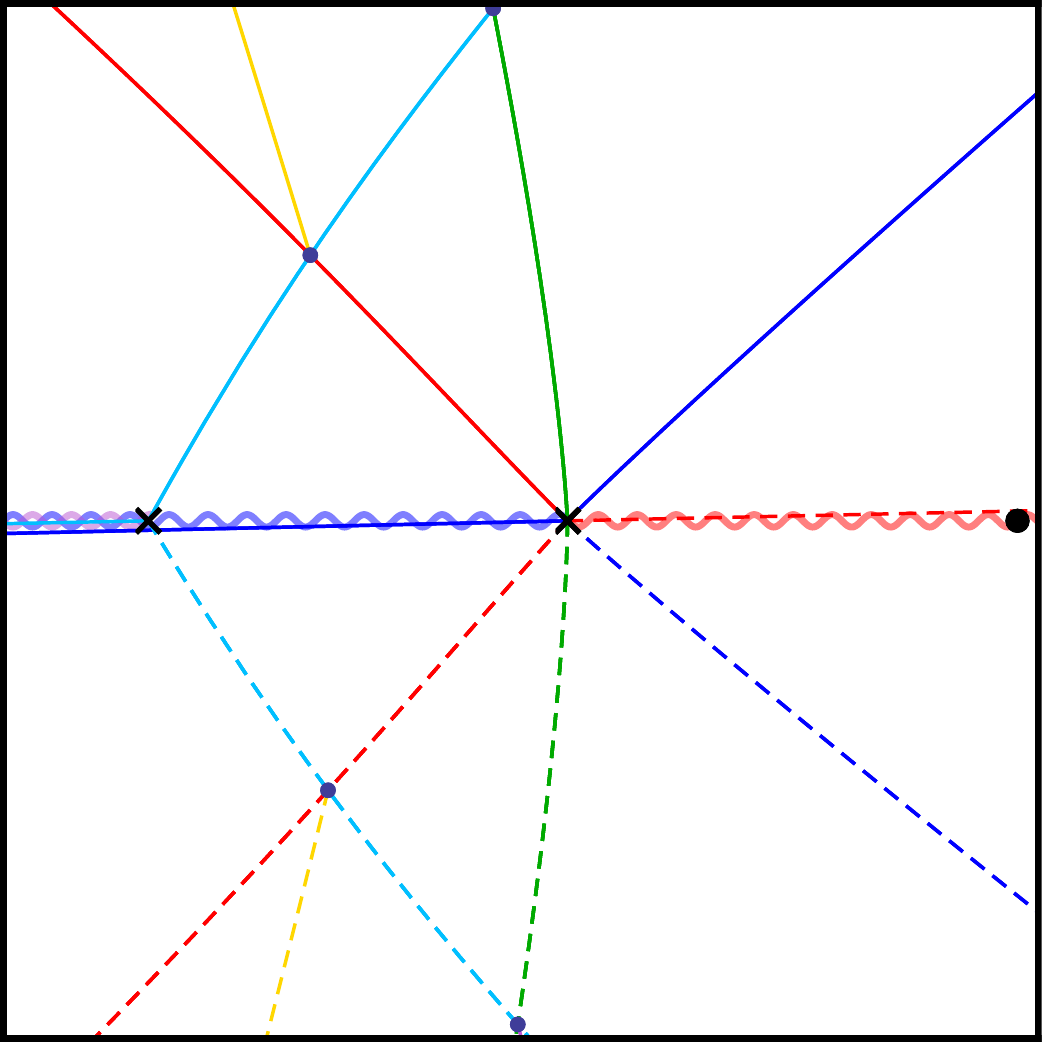}
		\caption{$\vartheta \approx \vartheta_{42}$}
		\label{fig:IRMN4k_SWall_04}
	\end{subfigure}
	\begin{subfigure}{.3\textwidth}
		\centering
		\includegraphics[width=\textwidth]{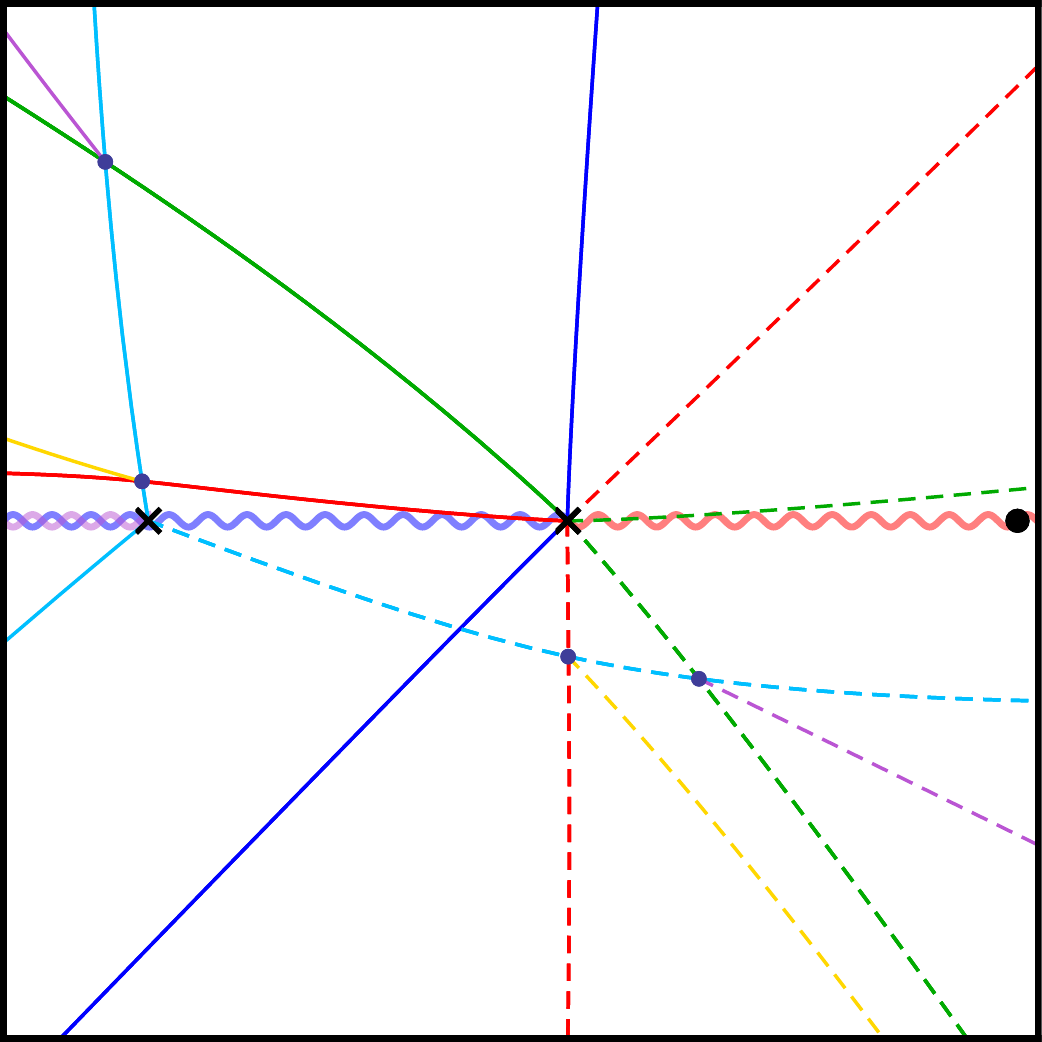}
		\caption{$\vartheta \approx \vartheta_{41}$}
		\label{fig:IRMN4k_SWall_05}
	\end{subfigure}
	\begin{subfigure}{.3\textwidth}
		\centering
		\includegraphics[width=\textwidth]{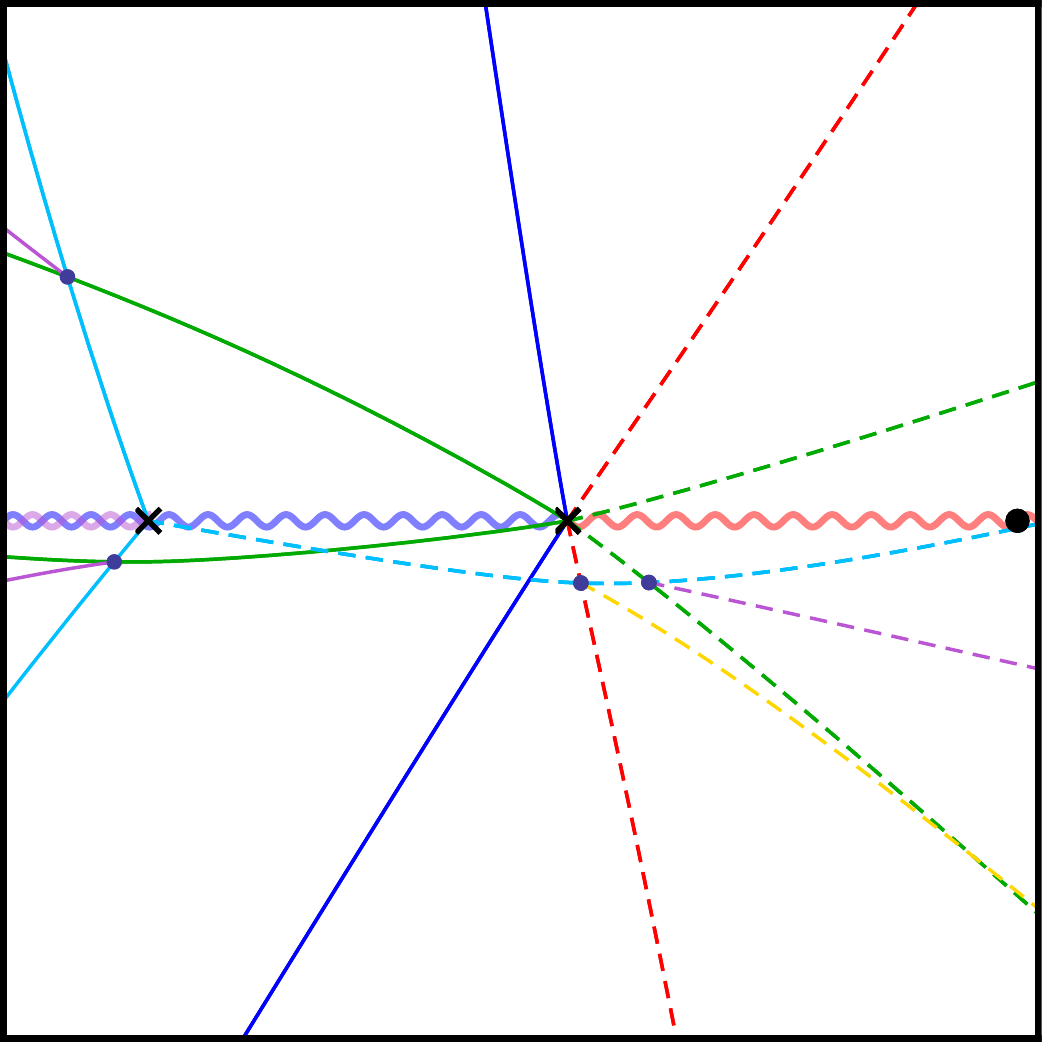}
		\caption{$\vartheta \approx \vartheta_{34}$}
		\label{fig:IRMN4k_SWall_06}
	\end{subfigure}	
	\caption{Rotation of the spectral network with $u_{2,3,4} \neq 0$.}
	\label{fig:IRMN4k_SWall_various_theta}
\end{figure}

Figure \ref{fig:IRMN4k_SWall_various_theta} shows the spectral network
at various values of $\vartheta$. At $\vartheta_{ij}$ there is
$\gamma_{ij}$ between one of the branch points and the endpoint of the
M2-brane, and Figure \ref{fig:IRMN4k_SWall_various_theta} is arranged
such that
\begin{align}
	0 < \vartheta_{13} < \vartheta_{43} < \vartheta_{12} < \vartheta_{42} < \vartheta_{41} < \vartheta_{34} < \pi.
\end{align}
That is, we can imagine the whole spectral network rotating
anti-clockwise as we increase $\vartheta$ from $0$ to $\pi$ and in the
course of the rotation we encounter six finite BPS strings. Therefore,
we can expect this theory to have twelve BPS states in total.  There
are four vacua, so there is one BPS state for each boundary condition
at the left and the right spatial infinity.

\begin{figure}[ht]
	\centering
	\begin{subfigure}{.3\textwidth}
		\centering
		\includegraphics[width=\textwidth]{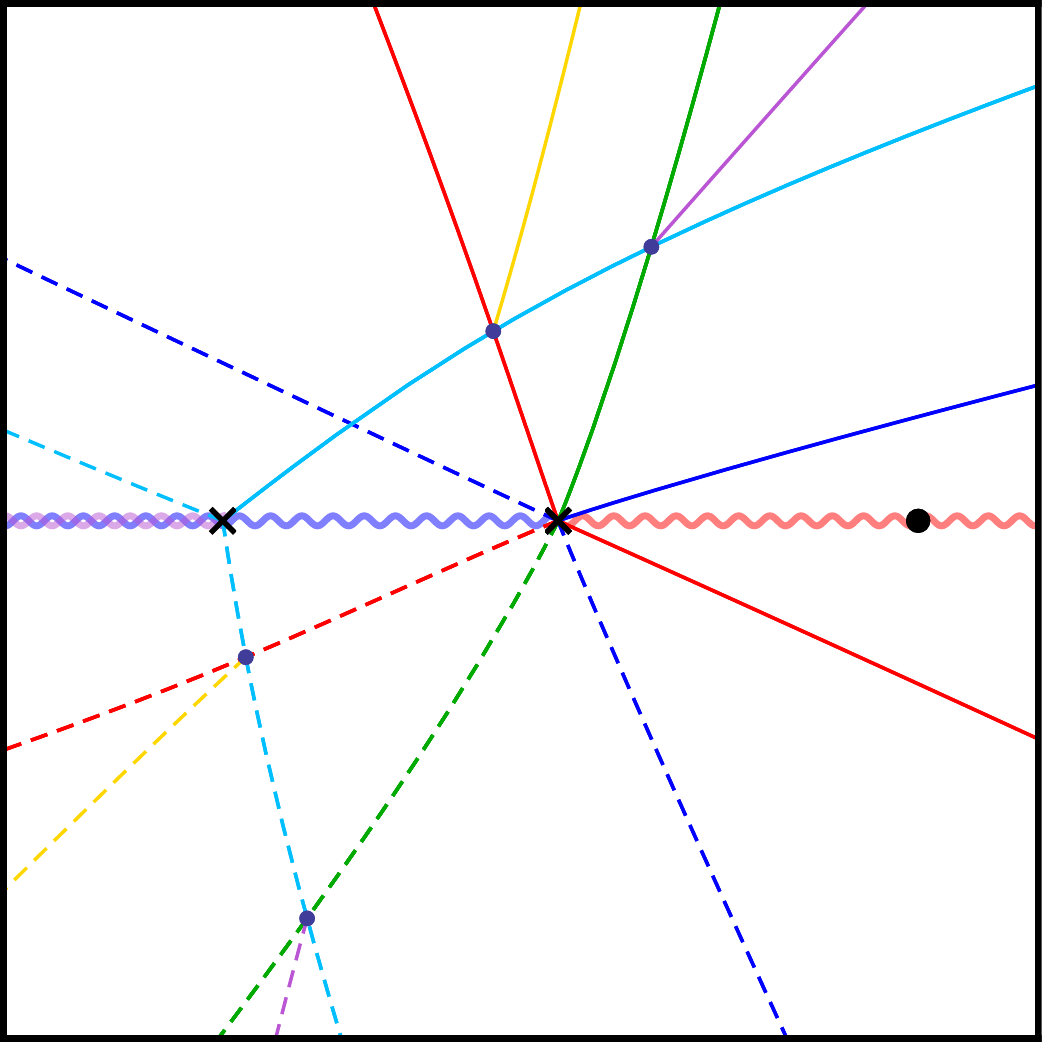}
		\caption{$| u_2| \sim 1$}
		\label{fig:IRMN4k_SWall_a_large}
	\end{subfigure}
	\begin{subfigure}{.3\textwidth}
		\centering
		\includegraphics[width=\textwidth]{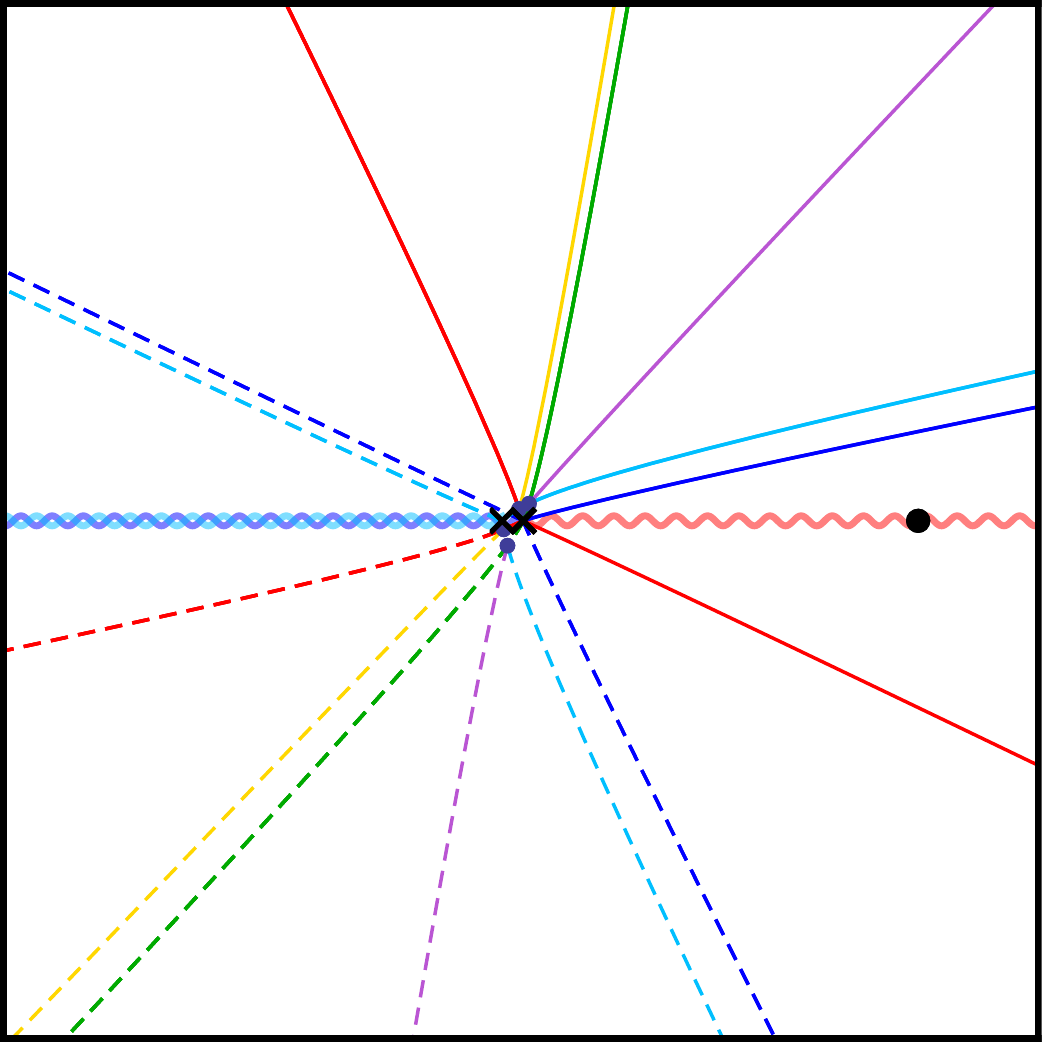}
		\caption{$| u_2| \ll 1$}
		\label{fig:IRMN4k_SWall_a_small}
	\end{subfigure}
	\begin{subfigure}{.3\textwidth}
		\centering
		\includegraphics[width=\textwidth]{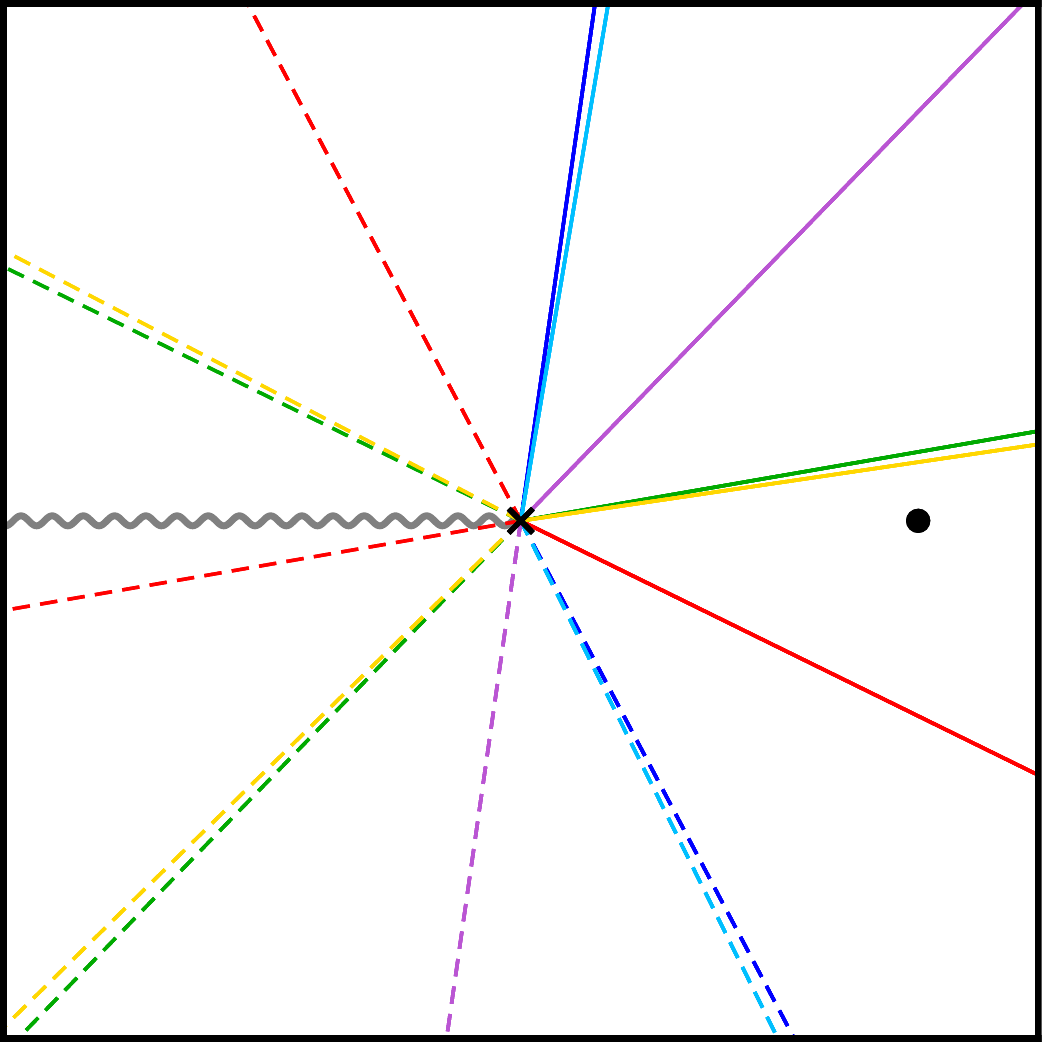}
		\caption{$ u_2 = 0$}
		\label{fig:IRMN4k_SWall_a_zero}
	\end{subfigure}
	
	\caption{Evolution of the spectral network under the limit of $u_{2,3} \to 0$.}
	\label{fig:IRMN4k_SWall_limit}
\end{figure}

Let us consider what happens when we have just one branch point of
ramification index 4. This limit corresponds to $ u_2,  u_3 \to 0$,
and the evolution of the spectral network under the limit is depicted
in Figure \ref{fig:IRMN4k_SWall_limit}.  Thus we see that the spectrum
of the BPS states at general $ u_{2,3,4}\neq 0$ is smoothly connected
to the more symmetric situation analyzed in Sec.~\ref{symmetric} with
$ u_{2,3}=0$ and $ u_4\neq 0$.


\section{\texorpdfstring{$S^2$}{S \textasciicircum 2} partition functions}\label{sec:pf}
As a final check of our proposal, we show in this section that the
partition function on $S^2$ of the 2d $\cN{=}(2,2)$ $\UU(k)$ gauge
theory with twisted superpotential
$\cW=\tr P(\Sigma)+\pi i (k+1){\rm tr}(\Sigma)$ in the infrared
limit agrees with that of the Landau-Ginzburg model with chiral fields
$X_1,\ldots, X_k$ with appropriately chosen superpotential
$W=W(X_1,\ldots,X_k)$.  We employ the localization methods recently
developed in \cite{Benini:2012ui,Doroud:2012xw,Gomis:2012wy}. 
The derivation can be easily generalized to arbitary gauge
group, and the quasihomogeneity of $P$ and $W$ is not required,
either.  The integrals below are only conditionally convergent. In
this section we perform the comparison of the partition functions
rather naively. The convergence issues will be explained in
Appendix~\ref{convergence}. It will be then clear that the
manipulations can be readily justified.

The partition function of the Landau-Ginzburg model of
$k$ variables $X_1,\ldots, X_k$ with the superpotential
$W(X_1,\ldots,X_k)$ is given by \cite{Gomis:2012wy} \begin{equation}
  Z_\text{LG}=(r\Lambda)^k
\int\limits_{\,\,\bC^k\!\!\!\!} \prod_a dX_a d\overline{X}_a
e^{-ir[W(X)+\overline{W}(\overline{X})]}.\label{LG-Z}
\end{equation}
where $r$ is the radius of the sphere.
The factor in front, $(r\Lambda)^k$, 
with $\Lambda$ being a renormalization scale, was not explicitly in
\cite{Benini:2012ui,Doroud:2012xw,Gomis:2012wy} but its presence is
mentioned in a footnote of \cite{Doroud:2012xw} and the computation
was done by the authors of these papers \cite{Jaume,Sungjay}.
The same applies to $(r\Lambda)^{-k^2}$ in (\ref{Zgaugegen}) below.
See \cite{Hori:2013ika} for a detailed explanation in a related context.
When  $W$ is quasi-homogeneous, a rescaling of fields
can absorb the $r$ in the integrand and yields the expected behaviour
$Z_{\rm LG}\sim r^{\hat{c}}$ with $\hat{c}$ being
the expected central charge of the infra-red fixed point
of the model \cite{Martinec:1988zu,Vafa:1988uu}.

The partition function of the $\cN{=}(2,2)$ supersymmetric gauge theory
was first computed in \cite{Benini:2012ui,Doroud:2012xw}
up to a sign factor which was later corrected in \cite{Honda:2013uca,Hori:2013ika}.
The one for the theory with
gauge group $\UU(k)$ and with the twisted
superpotential $\cW(\Sigma)$ is given by
 \begin{multline}
Z_\text{gauge}=(r\Lambda)^{-k^2}\sum_{m\in \bZ^k}
\int\limits_{\,\,\bR^k\!\!\!\!} \prod_a d(r\tau_a)
\prod_{a<b}\left(r^2(\tau_a-\tau_b)^2 + {(m_a-m_b)^2\over 4}\right) \\
\times (-1)^{(k+1)\sum_{a}m_a} 
e^{-i r[\cW(\Sigma) + \overline{\cW}(\overline{\Sigma})]}
\label{Zgaugegen}
\end{multline}
where \begin{equation}
  \Sigma=\diag(\tau_1,\ldots,\tau_k)+\frac{i}{2r}\diag(m_1,\ldots, m_k)
\end{equation} 
in the exponent.\footnote{The sign factor $(-1)^{(k+1)\sum_{a}m_a}$ was not in
\cite{Benini:2012ui,Doroud:2012xw}.
Its presence only changes the weight of
the sum over the topological type of the $\UU(k)$ gauge bundle, only when
$k$ is even. Therefore such a factor is rather sutble.
The presence  is demanded 
 for the factorization of the sphere partition function into 
two hemispheres \cite{Honda:2013uca,Hori:2013ika}.
As we will see, its presence is also needed for the match with the partition
function of the proposed Landau-Ginzburg model.}
For the twisted superpotential
$\cW(\Sigma)={\rm tr}P(\Sigma)+\pi i (k+1){\rm tr}(\Sigma)$,
the formula (\ref{Zgaugegen}) reads
\beq
Z_\text{gauge}=
\Lambda^{-k^2}\sum_{m\in \bZ^k}
\int\limits_{\,\,\bR^k\!\!\!\!} \prod_a d\tau_a
\prod_{a<b}\left((\tau_a-\tau_b)^2 
+ \left({m_a-m_b\over 2r}\right)^2\right)
e^{-i r[{\rm tr}P(\Sigma) + {\rm tr}\overline{P}(\overline{\Sigma})]}
\label{gauge-Z}
\eeq

Now, look at the infra-red regime $r\Lambda\gg 1$.
The sum $\sum_{m\in \bZ^k}$ in \eqref{gauge-Z} turns into an
integral $(2r)^k\int_{\bR^k} \prod_a d\upsilon_a$ for $\upsilon_a={m_a\over 2r}$,
and we have
\begin{equation}
Z_\text{gauge}\stackrel{r\Lambda\gg 1}{\longrightarrow}
\Lambda^{-k^2}r^k\int\limits_{\,\,\bC^k\!\!\!\!}  
\prod_a d\sigma_a d\overline{\sigma}_a \prod_{a<b} {|\sigma_a-\sigma_b|^2}
e^{-ir[\tr P(\Sigma)+\tr \overline{P}(\overline{\Sigma})]}\label{gauge-Z-IR}
\end{equation}  where $\sigma_a=\tau_a+i\upsilon_a$ and
\begin{equation}
\Sigma=\diag(\sigma_1,\ldots,\sigma_k).
\end{equation}

Let us introduce variables $X_a$ as the elementary symmetric polynomials of $\sigma_a$; equivalently, let us take
\begin{equation}
\det(z-\Sigma)=\sum_a X_a z^{k-a}. \label{sym}
\end{equation} 
where $z$ is a dummy variable. Then the Jacobian between the variables
$\sigma_a$ and the variables $X_a$ are given as in \eqref{jac}, \begin{equation}
 \det\left(\frac{\partial X_b}{\partial \sigma_a}\right)_{1\le a,b\le k}= \prod_{1\le a<b\le k} {(\sigma_a-\sigma_b)}.
\end{equation} 
Therefore, we see that the gauge partition function in the infrared,
\eqref{gauge-Z-IR}, agrees with the Landau-Ginzburg partition function
\eqref{LG-Z}, under the identification
\begin{equation}
P(\Sigma)=W(X_1,\ldots,X_k).
\end{equation}

We now have the equality of $S^2$ partition functions of the
$\UU(k)$ theory with the twisted superpotential in the infrared limit
and those of the Landau-Ginzburg theory.
Two-point functions of BPS operators can be
dealt with in the completely same way, by just inserting the operators
in the integral.
It is well-known that the resulting integral expressions suffer from
subtleties: apparently spurious operators do not decouple
and the choice of representatives of the (anti)chiral
ring elements matters \cite{Cecotti:1991me}.
The agreement holds provided that the operators
in the gauge system are identified with those in the Landau-Ginzburg model
precisely via the isomorphism
$\bC[\sigma_1,\ldots,\sigma_k]^{\mathfrak{S}_k}\,\cong\,
\bC[X_1,\ldots,X_k]$.


\acknowledgments

It is a pleasure for the authors to thank helpful discussions with Keshav Dasgupta, Nick Dorey, 
Sasha Getmanenko,
Jaume Gomis, Sangmin Lee, Sungjay Lee, Todor Milanov, Andy Neitzke,
Kyoji Saito,
John H.\ Schwarz,
Jaewon Song, Edward Witten, and Piljin Yi. C.~Y.~P.~would like to
thank Kavli IPMU for hospitality and support while this work was in the
initial and final stages. C.~Y.~P.~would also like to thank the
organizers of ``$\cN{=}2$ JAAZ'' Workshop at McGill University and the
10th Simons Summer Workshop in Mathematics and Physics for hospitality
and support where part of this work has been done. The work of
C.~Y.~P. is supported in part by Samsung Scholarship. The work of
K.~H. and
Y.~T. is supported in part by World Premier International Research
Center Initiative (WPI Initiative), MEXT, Japan through the Institute
for the Physics and Mathematics of the Universe, the University of
Tokyo. The work of K.~H. is also supported in part by
JSPS Grant-in-Aid for Scientific Research No. 21340109,
and the work of Y.~T. is supported in part by JSPS Grant-in-Aid for Scientific Research No. 25870159.

\appendix 
\section{On Kazama-Suzuki models and their Landau-Ginzburg descriptions}\label{KSreview}
The Kazama-Suzuki  model \cite{Kazama:1988qp} is a coset model 
 $G_k/H$ where $G$ is a compact simple simply connected Lie group,
$H$ is its closed subgroup of the same rank as $G$
such that the space $G/H$ of left cosets is K\"ahler;
 $k$ is a positive integer. It can be realized as a gauge theory
\cite{Witten:1991mk}:
the gauge group is $H/Z_G$ ($Z_G$ is the center of $G$)
and the matter theory is the direct product of
the $G_k$ Wess-Zumino-Witten model and the
$\mathfrak{g}/\mathfrak{h}$-valued free fermion,
where $H$ acts on $G$ and $\mathfrak{g}/\mathfrak{h}$ by the conjugation.
The models relevant for us are a subclass of
\begin{align}
\frac{\SU(m+n)_k}{\mathrm{S}[\UU(m)\times \UU(n)]}
\end{align} 
with the central charge
\begin{align}
	\hat{c} = \frac{c}{3} = \frac{kmn}{k+m+n}. 
\end{align}
This model is invariant under permulations of $k,m,n$ \cite{Kazama:1988qp}.
The model with $m=n=1$, i.e. $\SU(2)_k/\UU(1)$,
is equivalent to the ${\mathcal N}=2$ $A_{k}$ minimal model
\cite{Schwarz:1988qf}. The model with $m=1$, $n=N-k$, i.e.
\beq
{\SU(N)_1\over {\rm S}[\UU(k)\times \UU(N-k)]},
\eeq
is believed \cite{Lerche:1989uy, Gepner:1988wi} to be
equivalent to the IR fixed point of a Landau-Ginzburg model with a
superpotential $W(x_1,\ldots, x_k)$ which is chosen so that
\begin{align}
	 W(x_1,\ldots, x_n)=  \sum_{b=1}^{k} {\sigma_b}^N,
\end{align}
where $\sigma_b$ are auxiliary variables such that $x_b$ are their
elementary symmetric polynomials:
\begin{align}
	x_b = \sum_{1 \leq l_1 < l_2 < \cdots < l_b \leq k} \sigma_{l_1} \sigma_{l_2} \cdots \sigma_{l_b}.
\end{align}

One piece of evidence of the equivalence comes from computing the
central charge and the spectrum of the operators on each side and
matching them. In addition, when $k > N-k$, we can re-express
everything in terms of $N-k$ chiral fields, which implies $k
\leftrightarrow N-k$ duality \cite{Gepner:1988wi}. Another nontrivial
evidence comes from the calculation of elliptic genera in the two
descriptions, which yields agreement \cite{Witten:1993jg,DiFrancesco:1993dg}.

\section{Some algebra}\label{Morse}
We show that, when $W(x)$ is a Morse polynomial of $k$ variables,
$x=(x_1,\ldots, x_k)$,
a polynomial $\phi(x)$ that vanishes at all the critical points of $W(x)$
belongs to the ideal $I_W=(\partial_{x_1}W(x),\ldots,\partial_{x_k}W(x))$.
\footnote{We learned this proof from Kyoji Saito.}
 We have the following exact sequence
of sheaves of ${\mathcal O}$ modules on $\C^k$,
where ${\mathcal O}$ is the sheaf
of algebraic functions of $x_1,\ldots, x_k$:
\beq
0\to {\mathcal I}_W\to {\mathcal O}\to {\mathcal O}/{\mathcal I}_W\to 0.
\eeq
${\mathcal I}_{W}$ is the sheaf generated by the first derivatives of
$W(x)$. This yields an exact sequence of rings of global sections
\beq
0\to\Gamma(\C^k,{\mathcal I}_W)\to \C[x_1,\ldots,x_k]\to
\Gamma(\C^k,{\mathcal O}/{\mathcal I}_W)
\label{exact1}
\eeq
Since $W(x)$ is Morse, the derivatives
$\partial_{x_1}W(x),\ldots,\partial_{x_k}W(x)$ can be regarded as local
coordinates at each critical point $p$ of $W(x)$. 
Thus, $\phi(x)$, which vanishes at $p$, can be written as
$\sum_{a=1}^kg_a(x)\partial_{x_a}W(x)$
for some rational functions $g_a(x)$ which are regular in a neighborhood
of $p$. Therefore, the image of $\phi(x)$ in 
$\Gamma(\C^k,{\mathcal O}/{\mathcal I}_W)$ vanishes. By the exactness
of (\ref{exact1}), $\phi(x)$ should come from $\Gamma(\C^k,{\mathcal I}_W)$.
It remains to show $\Gamma(\C^k,{\mathcal I}_W)=I_W$, that is, any
global section of ${\mathcal I}_W$ can be written as
$\sum_{a=1}^kh_a(x)\partial_{x_a}W(x)$ for some polynomials
$h_1(x),\ldots,h_k(x)$.
For this, we consider another exact sequence of sheaves of
${\mathcal O}$-modules,
\beq
0\to {\mathcal K}\to {\mathcal O}^{\oplus k}\to {\mathcal I}_W\to 0
\eeq
where the right map is defined by $(s_1(x),\ldots,s_k(x))\mapsto
\sum_{a=1}^ks_a(x)\partial_{x_a}W(x)$ and ${\mathcal K}$ is defined
to be the kernel sheaf. This yields an exact sequence
\beq
\C[x_1,\ldots,x_k]^{\oplus k}\to \Gamma(\C^k,{\mathcal I}_W)\to
{\rm H}^1(\C^k,{\mathcal K})=0.
\eeq
This shows what we wanted.

\section{Convergence of integrals}\label{convergence}
Let us now discuss the convergence of the integral \eqref{LG-Z}. The
integrand is a pure phase. When $W(X)$ is a nontrivial function, the
phase will oscillate greatly at infinity, which should guarantee the
convergence. Here we analyze the issues of the convergence with more
care. 
\footnote{The authors learned the treatment presented here
from Alexander Getmanenko (a guidance including the reference
\cite{Shubin}), Yoshitsugu Takei (useful comment) and
Edward Witten (explicit instruction).}

Let us consider in general an oscillatory integral  
\begin{equation}
Z=\int_{\R^n} dx_1\cdots dx_n\,  \e^{iP(x_1,\ldots,x_n) }\label{conditional}
\end{equation} 
where $(x_1,\ldots,x_n)\in \bR^n$ and $P$ is a real function. We assume
that $|P'(x)|^2:=\sum_{j=1}^n|\partial_jP(x)|^2$
grows faster than a positive power of $|x|^2=\sum_{j=1}^n|x_j|^2$ at infinity:
there is some $\alpha>0$ and $C>0$
\beq
1+|P'(x)|^2\geq C(1+|x|^2)^{\alpha}\qquad \mbox{for any $x$}.
\label{growth}
\eeq

The right hand side of \eqref{conditional} is only conditionally convergent.
We would like to show that the absolutely convergent integral
\begin{equation}
Z_{\epsilon,f}=\int_{\R^n} dx_1\cdots dx_n \,
 \e^{ - \epsilon f(x_1,\ldots,x_n)+iP(x_1,\ldots,x_n)}
\end{equation} 
with a positive number $\epsilon$ and a positive function $f$ which
grows at least quadratically at infinity, has a limit when
$\epsilon\to 0$ independent of $f$. We define this limit to be the
left hand side of \eqref{conditional}
\begin{equation}
Z = \lim_{\epsilon \searrow +0} Z_{\epsilon,f}
\end{equation} 
independent of $f$.  With this interpretation of the integral, the
manipulation in Sec.~\ref{sec:pf} can be justified.

Let us introduce a differential operator
\beq
D:={1\over 1+|P'(x)|^2}\left(1-i\sum_{j=1}^n\partial_jP(x)\partial_j\right)
\eeq
and its formal adjoint
\beq
L:=\left(1+i\sum_{j=1}^n\partial_jP(x)\partial_j\right)
{1\over 1+|P'(x)|^2}\times.
\eeq
Using $D\e^{iP(x)}=\e^{iP(x)}$ we find
\beqa
Z_{\epsilon.f}&=&\int_{\R^n}\dd^n x\,
\e^{-\epsilon f(x)}\,D\e^{iP(x)}
=\int_{\R^n}\dd^n x\,L\!\left[\e^{-\epsilon f(x)}\right]\,\e^{iP(x)}
\nn\\[0.2cm]
&&\cdots\mbox{do it $N$ times}\cdots\nn\\[0.2cm]
&=&\int_{\R^n}\dd^nx\,L^N\!\left[
\e^{-\epsilon f(x)}\right]\,\e^{iP(x)}
\label{eqeqeq}
\eeqa
The partial integration is valid as long as $\epsilon>0$ due to the
exponential decay. One can show that $L^N\!\left[\e^{-\epsilon f(x)}\right]$
decays as fast as $1/|P'(x)|^N$  for any $\epsilon\geq 0$ including
$\epsilon=0$.
By the assumption (\ref{growth}), if we take $N$ such that $N\alpha>n$,
 the right hand side of (\ref{eqeqeq})
is absolutely convergent for any $\epsilon\geq 0$.
By the dominated convergence theorem, 
$Z_{\epsilon, f}$ has a limit as $\epsilon\searrow 0$ which is given by
\beq
\lim_{\epsilon\searrow 0}
Z_{0,f}=\int_{\R^n}\dd^n x\,L^N[1]\e^{iP(x)},
\eeq
for any $N$ such that $N\alpha>n$.
The result is obviously independent of $f$.
This was what we wanted to demonstrate.

\bibliographystyle{JHEP}
\bibliography{HPT}

\end{document}